\documentclass[11pt,a4paper]{article}

\usepackage[margin=0.95in]{geometry}
\usepackage[T1]{fontenc}
\usepackage[utf8]{inputenc}
\usepackage{lmodern}
\usepackage{microtype}
\usepackage{amsmath,amssymb}
\usepackage{graphicx}
\usepackage{booktabs}
\usepackage{tabularx}
\usepackage{multirow}
\usepackage{array}
\usepackage{xcolor}
\usepackage{xurl}
\usepackage[numbers,sort&compress]{natbib}
\usepackage[colorlinks=true,linkcolor=blue!60!black,citecolor=blue!60!black,urlcolor=blue!60!black]{hyperref}
\usepackage{caption}
\usepackage{float}
\usepackage{titlesec}
\usepackage{authblk}

\graphicspath{{figures/}}

\titleformat{\section}{\large\bfseries\sffamily}{\thesection}{0.8em}{}
\titleformat{\subsection}{\normalsize\bfseries\sffamily}{\thesubsection}{0.7em}{}
\titleformat{\subsubsection}{\normalsize\itshape}{\thesubsubsection}{0.6em}{}

\title{\Large\bfseries\sffamily Model Capacity and the Limits of Reconstruction Metrics in ECG and sEMG Denoising}

\author[1,*]{Jasmeet Singh Bindra}
\author[2]{Siddharth Panwar}
\affil[1]{Indian Knowledge Systems and Mental Health Applications (IKSMHA) Center, Indian Institute of Technology Mandi, Himachal Pradesh, India}
\affil[2]{School of Computing and Electrical Engineering, Indian Institute of Technology Mandi, Himachal Pradesh, India}
\affil[*]{Correspondence: \texttt{D24059@students.iitmandi.ac.in}; \texttt{siddharthpanwar@iitmandi.ac.in}}

\date{\small Preprint, November 2026}

\begin{document}
\maketitle

\begin{abstract}
Deep neural denoisers for electrocardiography (ECG) and surface electromyography (sEMG) vary widely in parameter count, yet evaluations rarely isolate model capacity or establish whether reconstruction metrics predict downstream utility. Holding a depthwise-separable convolutional backbone and training recipe fixed, we sweep capacity across 392--40,262 parameters for ECG and 392--153,718 for sEMG alongside controlled baselines, conditional diffusion, noise-pool and budget controls, and hierarchical downstream evaluation across seven clinical ECG and sEMG cohorts. On beat-based ECG, correlation peaks at 3.2K--6.5K parameters and declines at 40.3K, with nested noise-pool restrictions indicating sensitivity to limited training-noise diversity; conversely, continuous-window mixed-noise ECG improves monotonically to 40.3K parameters. On sEMG, correlation shows diminishing returns beyond 3.2K parameters, whereas signal-to-distortion ratio (SDR) and mean-frequency error improve to 153.7K parameters for power-line interference and exhibit an optimization confound for cardiac interference. Downstream ECG gains on unmodified clinical recordings are modest and classifier- and cohort-dependent (aiding deterministic pipelines on PTB-XL and CinC 2017, with no supported gain for \texttt{xresnet1d101} or CPSC2018). Under synthetic sEMG contamination at $-10$~dB signal-to-noise ratio (SNR), denoising improves gesture accuracy by 6.9--16.2 percentage points while satisfying non-inferiority on real GRABMyo recordings. For linear sEMG decoders, within-subject metric--utility slopes reverse from positive at $-10$ and $0$~dB to negative at $+10$~dB. Compact models achieve competitive performance against much larger baselines, but benchmark design and downstream tasks must guide evaluation.

\end{abstract}

\section{Introduction}

Deep neural networks are increasingly used to denoise ambulatory electrocardiography (ECG) and surface electromyography (sEMG) recordings, progressing from convolutional autoencoders and recurrent filters to multi-branch networks, diffusion models, Transformers, and state-space architectures \cite{antczak2018drnn,chiang2019fcndae,romero2021deepfilter,li2024descod,wang2025trustemgnet,morel2026drnetmamba,shih2026ssemgnet}. Parameter counts across these models span more than four orders of magnitude, from a few thousand to tens of millions. Despite this architectural growth, evaluation predominantly relies on clean-reference reconstruction metrics such as correlation coefficient (CC), root mean square error (RMSE), percent root-mean-square difference (PRD), and signal-to-distortion ratio (SDR) computed on semi-synthetic signal mixtures.

This evaluation convention leaves two central questions unresolved. First, how does denoising performance depend on model capacity when architecture, loss function, data splits, and optimization are held fixed? Published comparisons typically vary parameter count alongside architecture, objective, and training schedule, making it difficult to separate representational capacity from recipe differences or benchmark structure. This issue is particularly acute because widely used biosignal benchmarks draw noise from small, heavily reused pools: the standard QT Database (QTDB) plus MIT-BIH Noise Stress Test Database (NSTDB) ECG benchmark extracts all training and testing artifacts from three half-hour recordings \cite{laguna1997qtdb,moody1984nstdb,romero2021deepfilter}, and supervised sEMG benchmarks pair NinaPro DB2 signals with a small number of cardiac and motion records \cite{atzori2014ninapro,wang2023fcnemg,wang2025trustemgnet}. Second, do gains on clean-reference reconstruction metrics reliably predict better downstream performance on unmodified clinical ECG interpretation or myoelectric gesture control?

In our prior EEG denoising study \cite{bindra2026eeg}, sweeping channel width in a minimal depthwise-separable convolutional U-Net revealed early reconstruction saturation on muscular artifacts, modest continued gains on ocular artifacts, and a classifier-dependent metric--utility gap in which matched denoising degraded classical spatial-filter motor-imagery decoding and sleep staging. Because cortical EEG differs markedly from cardiac and neuromuscular signals, it remains open whether those patterns reflect EEG-specific dynamics or broader principles of biosignal denoising. ECG is quasi-periodic with stereotyped P--QRS--T morphology, whereas sEMG is a broadband stochastic interference pattern whose practical utility is commonly mediated by amplitude and spectral features rather than sample-wise phase.

To address these ECG- and sEMG-specific questions and facilitate comparison across modalities, we conduct a controlled investigation of capacity, benchmark design, and downstream utility. Our study makes three main contributions:
\begin{enumerate}
    \item \textbf{Controlled capacity sweeps across benchmarks, protocols, and training budgets.} Holding backbone and training recipe fixed across 392 to 40,262 parameters in ECG and 392 to 153,718 parameters in sEMG, we examine capacity effects alongside benchmark and optimization influences using conservative equivalence testing, nested noise-pool restrictions, continuous-window vs.\ beat-based ECG comparisons, source-cluster bootstraps, 75-epoch controls, and a second-family Patch-Transformer sweep.
    \item \textbf{Same-pipeline and published-recipe benchmarking with compute profiling.} We compare compact models against seven controlled neural baselines and published-recipe conditional diffusion (DeScoD-ECG), and profile parameter footprint and single-thread CPU execution latency.
    \item \textbf{Hierarchical downstream validation and metric--utility analysis across real and contaminated cohorts.} Treating patients, recordings, or subjects as the inferential unit, we evaluate downstream clinical ECG classification, rhythm, HRV, and delineation alongside within-subject sEMG gesture recognition on NinaPro DB2 and unseen real GRABMyo recordings, revealing a classifier-dependent utility gap in ECG and an SNR-dependent metric--utility sign reversal in linear sEMG decoding.
\end{enumerate}

\section{Related Work}

\subsection{Conventional and Deep Denoising for ECG and sEMG}

Classical biosignal denoising targets well-characterized spectral and morphological priors: zero-phase high-pass, FIR, or median filtering for ECG baseline wander and power-line interference \cite{vanalste1985ecg,dechazal2004heartbeats}, wavelet soft-thresholding for broadband muscle interference \cite{donoho1995wavelet}, and adaptive filtering or template subtraction for cardiac artifacts in sEMG \cite{lu2009emgecg}, with early QRS benchmark studies demonstrating how strongly detector performance depends on artifact spectra \cite{friesen1990qrs}. While computationally lightweight and interpretable, fixed filters struggle when artifacts overlap the physiological signal band, as occurs with electrode motion in ECG or cardiac interference below 50~Hz in sEMG.

Deep neural denoisers address band-overlapping artifacts via supervised regression on semi-synthetic mixtures. For ECG, most studies adopt the QTDB+NSTDB framework \cite{laguna1997qtdb,moody1984nstdb}, progressing from fully convolutional autoencoders (FCN-DAE) \cite{chiang2019fcndae} and recurrent networks (DRNN) \cite{antczak2018drnn} to Romero et al.'s DeepFilter Multibranch LANLD ($\approx$68.7K parameters) \cite{romero2021deepfilter}. Recent systems have scaled substantially: DeScoD-ECG applies conditional score-based diffusion with multi-shot averaging (1.93M parameters in released code; reported as 1.22M in the original paper) \cite{li2024descod}, TFCDiff employs 4.12M-parameter time-frequency diffusion \cite{li2025tfcdiff}, DR-net-Mamba introduces a 760K-parameter state-space backbone \cite{morel2026drnetmamba}, and EFIB-Net adds information-bottleneck attention \cite{ma2026efibnet}. At the compact end, Qiu et al.\ distilled a $\approx$6.9K-parameter ECG U-Net \cite{qiu2022lightweight}, and scaling studies of ECG classifiers from 20K to 200M parameters show flattening beyond $\approx 10^6$ parameters \cite{li2026ecgscaling}. Nevertheless, the independent effect of model capacity in ECG denoising remains insufficiently characterized under fixed recipes, and the influence of NSTDB's finite noise pool on capacity rankings has not been quantified.

For sEMG, supervised benchmarks commonly mix NinaPro DB2 recordings \cite{atzori2014ninapro} with cardiac signals from the MIT-BIH Normal Sinus Rhythm Database (NSRD), motion artifacts from NSTDB, and synthetic PLI. Following fully convolutional cardiac removal \cite{wang2023fcnemg}, TrustEMG-Net scaled to a 25.1M-parameter U-Net--Transformer hybrid \cite{wang2025trustemgnet}, and recent state-space models such as MSEMG \cite{liu2024msemg} and SSEMG-Net \cite{shih2026ssemgnet} further improved clean-reference metrics. Notably, Shih et al.\ observed that waveform-level gains do not automatically lower spectral frequency error \cite{shih2026ssemgnet}. Across these studies, model size is confounded with architecture family, and downstream myoelectric control is deferred to future work \cite{wang2025trustemgnet,liu2024msemg}.

\subsection{Benchmark Source Diversity and Task-Oriented Utility}

Scaling analyses treat parameter count as an explicit experimental axis \cite{kaplan2020scaling}, and equivalence testing distinguishes practical plateauing from underpowered null differences \cite{lakens2017equivalence}. Applying these tools to biosignal denoising exposes two methodological gaps. First, standard splits reuse a limited set of noise recordings across tens of thousands of segments. Even when recordings are split across time or channels to avoid direct window overlap \cite{li2025tfcdiff,morel2026drnetmamba}, the effect of limited noise-recording diversity on capacity curves and source-cluster uncertainty \cite{owen2007pigeonhole} remains largely unexamined. Second, task-oriented enhancement requires preserving features used by downstream clinical or myoelectric pipelines rather than minimizing sample-wise distance to a preprocessed target. Recent ECG papers have begun reporting downstream checks: DR-net-Mamba notes that reconstruction improvements on PTB-XL do not automatically transfer to classification \cite{morel2026drnetmamba}, and EFIB-Net pairs SNR/PRD with R-peak and beat-classification checks \cite{ma2026efibnet}. In sEMG, out-of-distribution benchmarks such as EMGBench \cite{yang2024emgbench} and multi-day corpora such as GRABMyo \cite{pradhan2022grabmyo} highlight session and electrode-placement shifts, yet deep sEMG denoisers have not been evaluated systematically for downstream gesture accuracy or same-condition metric--utility alignment.

\section{Methods}

\subsection{Experimental Design Overview}

Table~\ref{tab:design} summarizes all datasets, artifact sources, partitions, endpoints, and inferential units. Semi-synthetic mixtures evaluate capacity curves and benchmark-design sensitivity; unmodified clinical and myoelectric cohorts evaluate matched-pipeline downstream utility; and synthetically contaminated real-label cohorts estimate same-condition metric--utility slopes. Primary contrasts, equivalence margins, and statistical families were defined in prespecified analysis plans prior to execution, with documented secondary extensions noted in Appendix~\ref{app:provenance}.

\begin{table}[tbp]
\centering
\caption{Summary of experimental datasets, artifact sources, partitions, endpoints, and inferential units.}
\label{tab:design}
\scriptsize
\setlength{\tabcolsep}{3pt}
\begin{tabularx}{\textwidth}{@{} >{\raggedright\arraybackslash}p{2.5cm} >{\raggedright\arraybackslash}p{2.5cm} >{\raggedright\arraybackslash}p{3.0cm} >{\raggedright\arraybackslash}X >{\raggedright\arraybackslash}p{1.9cm} @{}}
\toprule
\textbf{Experiment / Cohort} & \textbf{Signal \& Artifact} & \textbf{Partition} & \textbf{Primary Endpoint(s)} & \textbf{Inferential Unit ($n$)} \\
\midrule
ECG beat-based recon.\ (QTDB+NSTDB) & Ch.\ 1 beats (360~Hz, 512 smp); NSTDB BW, EM, MA & 91 non-test recs (70/30 beat split: 50.4K/21.6K); 14 test recs (13.3K beats) & CC, SDR, T-RRMSE, S-RRMSE, SSD, PRD & Matched seeds (5); 2-way source \\
\addlinespace
ECG continuous recon.\ (QTDB+NSTDB) & Continuous 512-smp windows; uniform BW/EM/MA & 68 train / 23 val / 14 test QTDB recs (86.0K / 14.5K / 8.8K windows) & CC, SDR, T-RRMSE, S-RRMSE, SSD, PRD & Matched seeds (3) \\
\addlinespace
sEMG recon.\ (NinaPro DB2) & Active 2~s windows (1~kHz); NSRD ECG, NSTDB MOA, PLI & Subj.\ 1--25 train, 26--30 val, 31--40 test (18.1K test segs/task) & CC, SDR, $\text{SNR}_{\text{imp}}$, ARV-RMSE, MF-RMSE & Matched seeds (5); 2-way source \\
\addlinespace
PTB-XL v1.0.3 (unmodified) & 12-lead ECG (500~Hz), 5 superclasses & Folds 1--8 train (17.1K), fold 10 test (2,158 recs) & Patient macro AUROC (Det.\ LR; \texttt{xresnet1d101}) & Patient (1,877) \\
\addlinespace
CPSC2018 (unmodified) & 12-lead ECG (500~Hz), 9 classes & 5-fold stratified CV (out-of-fold) & Recording macro AUROC (Det.\ LR) & Recording (6,877) \\
\addlinespace
CinC 2017 (unmodified) & Single-lead ECG (300~Hz), 4 classes & 5-fold stratified CV (\texttt{REFERENCE-v3.csv}) & Challenge F1 (N/AF/Other); noisy share & Recording (8,528) \\
\addlinespace
MIT-BIH DS1/DS2 (unmodified \& noisy) & MLII ECG (360~Hz), AAMI beats & DS1 train (22 recs), DS2 test (22 recs; 15 unseen by QTDB$^*$) & R-peak F1, HRV SDNN/RMSSD err., beat acc. & Record (22; 15 unseen$^*$) \\
\addlinespace
QTDB delineation (unmodified \& noisy) & Ch.\ 1 ECG (250~Hz), \texttt{q1c} fiducials & 14 held-out QTDB test records & 4-fiducial composite MAE (3-fid.\ sens.) & Record (14) \\
\addlinespace
NinaPro DB2 gestures (clean \& noisy) & 12-ch.\ sEMG (1~kHz), 17 gestures + rest & Within-subj.\ reps 1/3/4/6 train, 2/5 test; cross-fold denoisers & Gesture accuracy (TD-LDA, TD-SVM, 1D-CNN) & Subject (40) \\
\addlinespace
GRABMyo v1.1.0 (unmodified real) & 16-ch.\ forearm / 12-ch.\ wrist sEMG (1~kHz), 17 classes & Within-session (trials 1--4/5--7) \& cross-day; 0-shot DB2 denoisers & Within-session \& cross-day TD-LDA acc.; LF/mains ratios & Subject (43) \\
\bottomrule
\end{tabularx}
\vspace{2pt}
\parbox{\textwidth}{\footnotesize $^*$See Section~\ref{sec:ecg_downstream} for MIT-BIH parent-record provenance: 7 of 22 DS2 records share parent recordings with QTDB training/validation excerpts; all MIT-BIH conclusions are verified on the 15 strictly unseen DS2 records.}
\end{table}

\subsection{Controlled Backbone, Signal Normalization, and Training Objective}

The denoising backbone follows the clean-only depthwise-separable convolutional \cite{chollet2017xception} U-Net with efficient channel attention (ECA) \cite{wang2020ecanet} introduced in \cite{bindra2026eeg}, varying only the base channel width $C$ while keeping depth, kernels, dilations, activations, normalization, loss, and optimization identical. In the benchmark packs, each noisy input segment $y \in \mathbb{R}^{1 \times T}$ is standardized by its segment standard deviation $\sigma_y = \mathrm{std}(y)$ (with a $+10^{-8}$ numerical guard during continuous downstream inference), yielding $\tilde{y} = y / \sigma_y$ and target $\tilde{x} = x / \sigma_y$. The network maps $\tilde{y} \mapsto \hat{\tilde{x}}$ through a $1\times 1$ stem ($1 \to C$), two encoder DSConv blocks ($k=9$, dilations $1, 2$) separated by stride-2 downsampling ($C \to 2C \to 4C$), three bottleneck DSConv blocks at width $4C$ (dilations $4, 8, 16$), two symmetric transposed-convolution decoder stages with skip concatenations, and a $1\times 1$ output head. Each DSConv block uses batch normalization, SiLU activation, and ECA ($k=3$), with residual connections where channel widths match.

Parameter count depends only on $C$ and is invariant to sampling rate and segment length: $C \in \{1, 2, 3, 4, 5, 6, \allowbreak 8, 10, 12, 16, 24, 32\}$ spans 392; 1,048; 1,990; 3,218; 4,732; 6,532; 10,990; 16,592; 23,338; 40,262; 87,838; and 153,718 trainable parameters. The primary sweep uses $C \in \{2, 4, 6, 8, 16\}$ (5 matched seeds, 42--46), supplemented by $\{1, 3, 5\}$ (5 seeds) around the elbow, $\{10, 12\}$ (3 seeds) for interpolation, and $\{24, 32\}$ (3 seeds, sEMG only) for the upper regime. Empirical gradient-support profiling yields a convolutional receptive field of 990 samples (plus global segment pooling via ECA), covering an entire 512-sample ECG segment (1.42~s at 360~Hz) and 0.99~s locally within a 2,000-sample sEMG segment (2.0~s at 1~kHz).

All controlled DSConv widths and all controlled neural baselines (ECG and sEMG) are trained through the identical engine with AdamW (learning rate $10^{-3}$, weight decay $10^{-4}$, gradient norm clipping at 1.0, batch size 256, 25 epochs with 3-epoch linear warm-up and cosine decay) and checkpoint selection by validation SDR. The objective is an SNR-weighted Charbonnier loss on the normalized clean estimate:
\begin{equation}
\mathcal{L}(\hat{\tilde{x}}, \tilde{x}) = \frac{1}{NT}\sum_{i=1}^{N}\sum_{t=1}^{T} w(\mathrm{SNR}_i)\,\sqrt{(\hat{\tilde{x}}_{i,t} - \tilde{x}_{i,t})^2 + \varepsilon^2},
\label{eq:loss}
\end{equation}
where $\varepsilon = 2\times 10^{-3}$ and $w(\mathrm{SNR}) = \mathrm{clip}\!\left((2 - \mathrm{SNR})/14,\, 0.25,\, 2.0\right)$ upweights lower-SNR segments using the realized input SNR in dB. Optimization controls retrain base4, base6, and base16 for 75 epochs (seeds 42--43) across all ECG and sEMG tasks, and retrain ECG base16 at learning rate $3\times 10^{-4}$.

\subsection{Semi-Synthetic Signal Mixture Models and Benchmarks}

All semi-synthetic mixtures follow an additive model $y = x + \alpha\,a$, where $x$ is a clean reference segment, $a$ is a noise segment, and $\alpha > 0$ sets the contamination level according to the modality's established benchmark convention.

\subsubsection{ECG benchmarks (QTDB + MIT-BIH NSTDB)}
For the beat-based benchmark, we replicate the DeepFilter/DeScoD-ECG builder \cite{romero2021deepfilter,li2024descod} on QTDB v1.0.0 (105 records at 250~Hz) \cite{laguna1997qtdb,goldberger2000physionet} and MIT-BIH NSTDB v1.0.0 (records bw, em, ma; 2 channels $\times$ 650,000 samples at 360~Hz) \cite{moody1984nstdb}. Channel-1 beats delimited by \texttt{pu1} annotations (from P-onset $-40$~ms to the next P-onset, dropping segments with $\ge 2$ normal QRS marks) are mirror-padded, resampled from 250 to 360~Hz via anti-aliased polyphase resampling (\texttt{scipy.signal.resample\_poly}), mean-endpoint-centered by subtracting $(b_0 + b_{\text{end}})/2$, and placed at offset 16 inside a 512-sample zero-padded vector (discarding beats $>496$ samples). This yields 72,002 beats across 91 non-test records and 13,316 test beats from the 14 fixed DeepFilter test records. Following the DeepFilter partitioning protocol, validation uses a random 70/30 beat-level split of the 91 non-test records (50,401 train / 21,601 validation beats), so train and validation share recordings whereas the 14 test records are completely disjoint.

Each NSTDB noise channel is split into two disjoint 15-minute halves of 634 non-overlapping 512-sample windows. Consecutive noise windows wrap cyclically across beats (mean reuse 79.5 in train, 34.1 in validation, 21.0 in test) and are scaled by peak-to-peak amplitude ratio $r \sim \mathcal{U}\{0.20, 0.21, \dots, 1.99\}$:
\begin{equation}
\alpha = r\,\frac{\max(x) - \min(x)}{\max(a) - \min(a)}.
\label{eq:ecg_mix}
\end{equation}
Baseline wander (BW) matches the published split; electrode motion (EM) and muscle artifact (MA) apply Eq.~\eqref{eq:ecg_mix} identically as protocol extensions. Noise version 1 (nv1: channel 1 first half train / channel 2 second half test) is primary; noise version 2 (nv2: channels swapped, 3 seeds) provides replication and pooled nv1+nv2 literature comparison. To examine sensitivity to training-noise-pool size via a nested noise-pool restriction, we also train base4, base6, and base16 (3 seeds) using only the first 25\% (158 windows) or 50\% (317 windows) of the training noise half while keeping test noise unchanged. Note that non-overlapping windows drawn from the same 15-minute recording are temporally correlated rather than independent recording sessions.

For continuous clinical evaluation, we build a continuous-window mixed-noise QTDB+NSTDB benchmark from the same 91 non-test and 14 test records, resampled continuously to 360~Hz (polyphase 36/25) into 512-sample zero-mean windows (no zero-padding; train stride 256, val/test stride 512). Validation is strictly record-level (68 train / 23 val records; 85,952 / 14,536 windows; 14 test records / 8,848 windows). Each window independently draws one artifact type uniformly from \{BW, EM, MA\} (not simultaneous superpositions) from the nv1 halves (1,902 distinct noise windows; mean train reuse 45.2), scaled via Eq.~\eqref{eq:ecg_mix} (mean input SNR $-8.35$~dB vs.\ $-14.3$~dB in zero-padded beats).

\subsubsection{sEMG benchmark (NinaPro DB2 + NSRD / NSTDB / PLI)}
Following TrustEMG-Net \cite{wang2025trustemgnet}, NinaPro DB2 sEMG recordings (40 subjects, 12 channels, 2~kHz) \cite{atzori2014ninapro} are band-pass filtered (20--500~Hz), resampled to 1~kHz, normalized per recording by $\max(|x|)$, and cut into non-overlapping 2~s (2,000-sample) windows retaining segments with $\ge 50\%$ active movement samples. Training uses subjects 1--25 (Exercise 1, channel 2), validation subjects 26--30, and testing subjects 31--40 (Exercise 2, channel 11; 3,610 clean segments $\times$ 5 test SNRs $= 18,050$ test segments per task). For target input SNR $s$ (in dB), noise scaling follows signal power:
\begin{equation}
\alpha = \sqrt{\frac{\|x\|_2^2}{(\|a\|_2^2 + 10^{-20})\,10^{s/10}}}.
\label{eq:emg_mix}
\end{equation}
Training SNRs are $\{1, -3, -7, -11, -15\}$~dB and test SNRs are $\{2, -2, -6, -10, -14\}$~dB across three single-contaminant tasks: (i) cardiac interference (ecg) from MIT-BIH NSRD (14 train records; 4 held-out test records: 19090, 19093, 19140, 19830); (ii) motion artifact (moa) from NSTDB em channel 1 (train) and channel 2 (test; disjoint channels from shared recordings, inherited from \cite{wang2025trustemgnet}); and (iii) drifting sinusoidal power-line interference (pli, 58.4--61.5~Hz) with disjoint train/test frequency grids.

\subsection{Controlled Baselines and High-Capacity References}

Under the identical splits, loss (Eq.~\ref{eq:loss}), and 25-epoch schedule (3 seeds, 42--44; architecture-only comparison), we train a 33-tap learned FIR (34 parameters), tiny TCN (365), Patch-Transformer sweep ($d_{\text{model}} \in \{8, 16, 24, 32, 48\}$, 1,345--40,705 on ECG; $\{24, 32\}$ on sEMG), and PyTorch ports of DRNN (25,537) \cite{antczak2018drnn}, DeepFilter LANLD (68,719) \cite{romero2021deepfilter}, FCN-DAE (80,223, ECG only due to divisible-by-32 pooling) \cite{chiang2019fcndae}, and the 8.46M EEGDenoiseNet CNN \cite{zhang2021eegdenoisenet} (with lr $3\times 10^{-4}$ and 75-epoch controls). For high-capacity ECG reference on nv1 BW, we evaluate official released weights and 3-seed retrained runs of DeScoD-ECG (1,926,321 parameters, 400 epochs, L1 diffusion loss, 1- and 10-shot) \cite{li2024descod}. Published sEMG figures for TrustEMG-Net (25.1M), U-Net (16.7M), FCN (11.2M), and CNN (2.7M) \cite{wang2025trustemgnet} provide approximately protocol-matched literature context: their evaluation used 3,180 clean test segments (vs.\ our 3,610), 4-fold cross-validation, combined 3- and 5-contaminant mixtures alongside single-artifact cases, extra MOA sources, and a different training objective.

Compute footprint is profiled on an idle AMD EPYC 7313 16-core CPU (single-thread PyTorch, batch size 1), with ONNX export parity verified for base4, base6, and base16 checkpoints.

\subsection{Downstream Clinical ECG and Myoelectric sEMG Evaluation}
\label{sec:ecg_downstream}

For continuous arbitrary-length recordings, each channel is resampled to the denoiser rate (360~Hz for ECG, 1,000~Hz for sEMG) via polyphase resampling (\texttt{resample\_poly}), segmented into overlapping windows (512 samples at stride 256 for ECG; 2,000 at stride 1,000 for sEMG, with the final window aligned to the signal end and signals shorter than one window edge-padded then trimmed), divided per window by its standard deviation ($\mathrm{std}(y_w) + 10^{-8}$, without per-window mean subtraction during overlap-add so physical baseline structure is preserved prior to downstream clinical filters), passed through the denoiser, \emph{rescaled back to physical amplitude by multiplying by the same factor}, combined via uniform overlap-add averaging, and resampled back to the native sampling rate.

All downstream evaluations test \emph{matched-input adaptation}: classifiers are trained and evaluated on the same preprocessing condition (\texttt{denoised/denoised} minus \texttt{raw/raw}). This assesses whether a downstream system benefits when adapted to denoised signals, rather than drop-in deployment into a fixed raw-trained classifier.

\subsubsection{Unmodified clinical ECG cohorts and source-record provenance}
We apply the continuous-window mixed-noise ECG denoisers (base2, base4, base6, base8, base16; 3 seeds) to five cohorts:
\begin{itemize}
    \item \textbf{PTB-XL v1.0.3} \cite{wagner2020ptbxl} (5 superclasses, 12 leads, 500~Hz, folds 1--8 train / 10 test: 17,084 / 2,158 records, 1,877 test patients; fold 9 unused in deterministic LR and monitored without early stopping in \texttt{xresnet1d101}; strictly external): (i) deterministic lead-II NeuroKit2 \cite{makowski2021neurokit2} RR/HRV plus per-lead median-beat morphology and Welch powers (0.5--5, 5--15, 15--40~Hz), median imputation, \texttt{StandardScaler}, and one-vs-rest class-balanced LR; (ii) neural \texttt{xresnet1d101} \cite{strodthoff2021ptbxl} (100~Hz polyphase, 2.5~s crops, BCE, AdamW wd $10^{-2}$, one-cycle max lr $10^{-2}$, 50 epochs, max pooling over 2.5~s test windows at 1.25~s stride; 3 paired seeds). Endpoint: patient macro AUROC.
    \item \textbf{CPSC2018} \cite{liu2018cpsc} (6,877 12-lead 500~Hz recordings, 9 classes; strictly external): same deterministic 12-lead pipeline via 5-fold CV stratified on the first listed SNOMED-CT label. Endpoint: recording macro AUROC.
    \item \textbf{PhysioNet/CinC 2017} \cite{clifford2017cinc} (8,528 single-lead 300~Hz recordings; Normal/AF/Other/Noisy; \texttt{REFERENCE-v3.csv} primary, \texttt{REFERENCE.csv} sensitivity; strictly external): 24 NeuroKit2 RR, template, and Welch features with median imputation, \texttt{StandardScaler}, and balanced multinomial LR in 5-fold stratified CV. Endpoints: challenge F1 and Noisy share.
    \item \textbf{MIT-BIH Arrhythmia DS1/DS2} \cite{moody2001mitbih,dechazal2004heartbeats} (MLII, 360~Hz; 22 DS1 / 22 DS2 records): Pan--Tompkins \cite{pan1985qrs} R-peak F1 ($\pm 150$~ms), HRV SDNN/RMSSD absolute error (Kubios-corrected 0.3--2.0~s intervals; N--N secondary), and AAMI beat accuracy (N/S/V/F; 200/600~ms median baseline removal, 4 RR + 19 QRS/T samples, \texttt{StandardScaler}, shrinkage LDA). \emph{Provenance:} Because 15 QTDB records derive from MIT-BIH \cite{laguna1997qtdb}, 10 entered continuous training and 3 validation (Appendix~\ref{app:provenance}), overlapping 7 DS2 records (100, 103, 117, 213, 221, 231, 232) and 4 DS1 records (114, 116, 223, 230). We prioritize the provenance-restricted 15 unseen-parent DS2 records (plus $n=14$ excluding record 202, shared with DS1 201) alongside full 22-record DS2.
    \item \textbf{QTDB delineation} (14 held-out test records): NeuroKit2 DWT (\texttt{ecg\_delineate}) vs.\ \texttt{q1c} annotations (P-onset, QRS-onset, QRS-offset, T-offset, $\pm 150$~ms). Primary: 4-fiducial MAE (3-fiducial excluding mislocated QRS-onset as sensitivity).
\end{itemize}

\subsubsection{Cross-fold sEMG denoisers and within-subject gesture recognition}
Mixed-contaminant sEMG denoisers train across all 12 NinaPro DB2 Exercise 1 channels by sampling one contaminant uniformly per 2,000-sample segment from \{ecg, moa, pli, baseline wander, white Gaussian noise\} (single contaminant per window) at an SNR drawn uniformly from $\{1, -3, -7, -11, -15\}$~dB in two subject-disjoint folds (Fold A: 1--20; Fold B: 21--40; widths base2, base4, base6, base8, base16; 3 seeds). Downstream gesture classifiers operate in the \emph{within-subject} regime (distinct from subject-independent classification):
\begin{itemize}
    \item \textbf{NinaPro DB2} (40 subjects, Exercise 1/B, 17 gestures + rest; train reps 1, 3, 4, 6; test 2, 5): clean and held-out test-noise mixtures ($-10, 0, +10$~dB), evaluated with Hudgins TD features \cite{hudgins1993myoelectric} (MAV, WL, ZC, SSC; threshold $0.01\times$ train SD; 200~ms windows, 100~ms step) using shrinkage LDA and linear SVM ($C=1$), plus a per-subject 1D-CNN (32/64/64 conv blocks, kernels 7/5/3, BN, ReLU, pooling, Adam lr $10^{-3}$, 20 epochs, 3 paired seeds) at base6 and base16 on clean, $-10$, and $0$~dB.
    \item \textbf{GRABMyo v1.1.0 real recordings} \cite{pradhan2022grabmyo} (43 subjects unseen in denoiser training $\times$ 3 days, 17 classes, 7 trials $\times$ 5~s; 16 forearm / 12 wrist channels; polyphase 2,048$\to$1,000~Hz, per-trial mean removed; no added noise): compares raw, conventional filtering (4th-order 20--450~Hz Butterworth + $Q=30$ notches at 60~Hz harmonics), and zero-shot DB2 denoisers (all 6 checkpoints per width evaluated independently and accuracies averaged across checkpoints and sessions). Within-session (trials 1--4/5--7) is primary and cross-day (Session 1 vs.\ 2--3) secondary, plus Welch $<20$~Hz and mains power ratios.
\end{itemize}

\subsection{Statistical Inference and Metric--Utility Analysis}

\textbf{Capacity equivalence.} Primary reconstruction metrics are CC, temporal RRMSE (T-RRMSE), spectral RRMSE (S-RRMSE via Welch PSD), and SDR, supplemented by ECG SSD, MAD, PRD, cosine similarity, and SNR improvement ($\text{SNR}_{\text{imp}}$), and sEMG average-rectified-value RMSE (ARV-RMSE) and mean-frequency RMSE (MF-RMSE; STFT magnitude mean frequency over 10--500~Hz in non-overlapping 200~ms windows over active-movement segments containing $\ge 50\%$ active samples, matching \cite{wang2025trustemgnet}). Full mathematical definitions (including the DeepFilter vs.\ standard PRD denominator conventions) are provided in Appendix~\ref{app:provenance}. We specify practical equivalence margins of $\delta = 0.005$ for CC/T-RRMSE/S-RRMSE and $\delta = 0.10$~dB for SDR. In ECG and sEMG denoising, 0.005 CC and 0.10~dB SDR sit well below run-to-run variation and are orders of magnitude smaller than differences across competitive architectures or noise levels, while remaining consistent with \cite{bindra2026eeg}. Each adjacent-width increment $\Delta$ (oriented so positive indicates improvement for the wider model) is summarized by a matched-seed paired-$t$ 95\% CI and critical margin $m^* = \max(|\text{CI}_{\text{lo}}|, |\text{CI}_{\text{hi}}|)$. Requiring the full 95\% CI to lie inside $[-\delta, +\delta]$ is strictly more conservative than conventional TOST at $\alpha = 0.05$ (which uses a 90\% CI). Contrasts are classified into four mutually exclusive categories: \emph{Equivalence} (95\% CI inside $[-\delta, +\delta]$), \emph{Positive gain} ($\text{CI}_{\text{lo}} > 0$ and outside $[-\delta, +\delta]$), \emph{Decrease} ($\text{CI}_{\text{hi}} < 0$ and outside $[-\delta, +\delta]$), or \emph{Inconclusive}. Primary contrasts are CC base6$\to$8 and base8$\to$16 per task (Bonferroni 99.17\% CIs checked as sensitivity); formal equivalence claims apply strictly to the tested contrasts rather than across arbitrary multi-step capacity spans. Source sampling uncertainty is estimated via 20,000-resample two-way pigeonhole bootstraps \cite{owen2007pigeonhole} crossing test records/subjects and noise windows; variance-added seed-plus-source intervals combine seed and source SEs in quadrature ($\mathrm{SE}_{\text{tot}} = \sqrt{\mathrm{SE}_{\text{seed}}^2 + \mathrm{SE}_{\text{source}}^2}$) under an independence approximation.

\textbf{Downstream and metric--utility inference.} Downstream tests treat the patient, recording (CinC 2017), or subject as the independent unit, averaging denoiser and classifier seeds within unit first. Primary tests use two-sided paired $t$-tests (or 10,000-resample unit bootstraps for pooled macro AUROC/F1) with 95\% unit bootstrap CIs, Wilcoxon sensitivity, and Benjamini--Hochberg FDR (BH-FDR) \cite{benjamini1995fdr} (plus Holm sensitivity \cite{holm1979}) within each family: primary clinical ECG (7 endpoints $\times$ 2 widths $= 14$ tests), PTB-XL neural and CPSC2018 replications (2 widths each), NinaPro DB2 primary TD-LDA (4 input states $\times$ 2 widths $= 8$ tests; TD-SVM 8 tests and 1D-CNN 6 tests as separate secondary families), and GRABMyo primary within-session TD-LDA (2 placements $\times$ 2 contrasts $\times$ 2 widths $= 8$ tests). GRABMyo non-inferiority vs.\ raw uses a prespecified practical tolerance of one percentage point ($-1$~pp), evaluated strictly by whether the lower 95\% CI bound exceeds $-1$~pp rather than by superiority $q$-values.

Same-condition metric--utility regressions model downstream change against oriented reconstruction metrics with unit fixed effects across 30 ECG tests (MIT-BIH R-peak F1 and QTDB delineation MAE; consecutive 512-sample blocks contaminated separately with held-out nv1 BW, EM, or MA test noise via Eq.~\eqref{eq:ecg_mix} across 3 contamination seeds with random noise-half offsets $\times$ 15 checkpoints; 3 noises $\times$ 5 metrics) and 24 sEMG tests (DB2 TD-LDA and TD-SVM, $n=40$; 3 SNRs $\times$ 4 metrics across 15 checkpoints). Primary inference uses CR1 cluster-robust SEs ($t_{G-1}$) with 95\% CIs, verified against CR2/Satterthwaite \cite{pustejovsky2018small}, Webb wild cluster bootstrap \cite{cameron2008bootstrap,webb2023reworking}, and two-stage unit-slope $t$ and Wilcoxon tests under separate BH-FDR corrections.

\section{Results}

\subsection{ECG Capacity Curves: Beat-Based Decline vs.\ Continuous-Window Gain}

On the standard beat-based QTDB+NSTDB benchmark (nv1, 5 seeds), reconstruction rises sharply from base1 (392 parameters, CC 0.577--0.685) to a CC peak at 3.2K--6.5K parameters (base4--base6), then declines at base16 (40.3K parameters) across all three artifact tasks (Figure~\ref{fig:ecg_cap}a--b; full fine-grid values in Appendix~\ref{app:capacity}, Table~\ref{tab:app_ecg_fine}). Peak test CC occurs at base5 for BW (0.929 $\pm$ 0.001, SDR 9.42~dB), base4--base5 for EM (0.816 $\pm$ 0.003, SDR 5.41~dB), and base6 for MA (0.928 $\pm$ 0.002, SDR 9.38~dB); BW and MA SDR reach shallow maxima at base10 (9.48 and 9.42~dB, within 0.06~dB of base5--base6) before dropping at base16.

\begin{figure}[tbp]
\centering
\includegraphics[width=\textwidth]{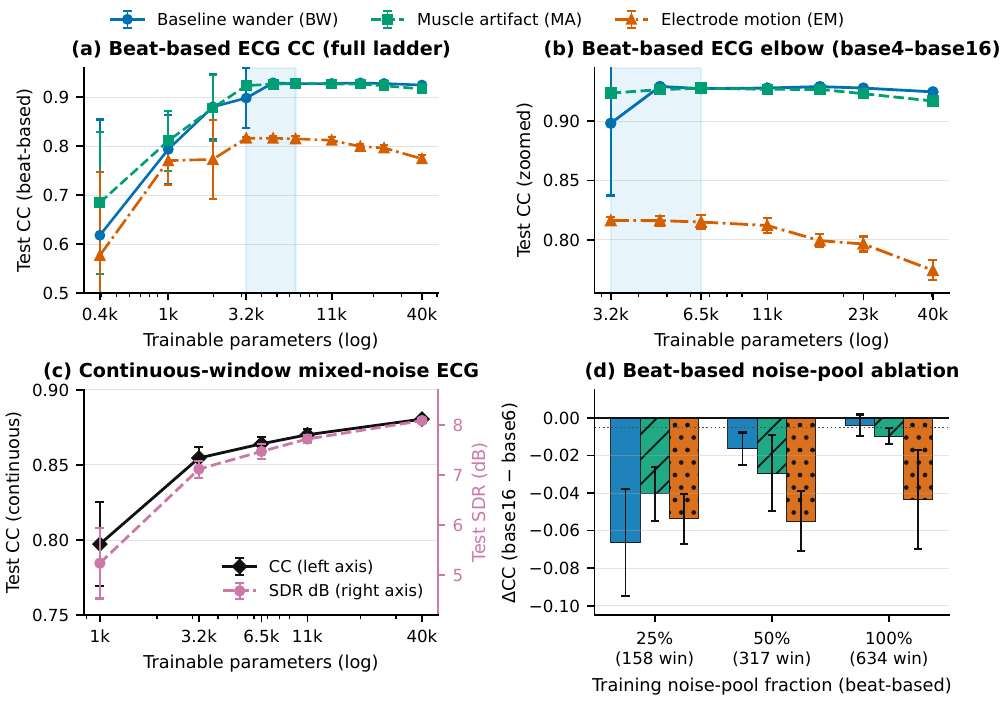}
\caption{Controlled ECG capacity curves and nested noise-pool restriction on QTDB+NSTDB (nv1). (a) Full beat-based ECG CC ladder (392--40,262 parameters, mean $\pm$ SD over 5 seeds; 3 seeds for base10/base12). (b) Zoomed view of the beat-based capacity elbow (base4--base16), showing peaks at 3.2K--6.5K parameters and degradation at 40.3K. (c) Continuous-window mixed-noise ECG reconstruction on the 14 held-out QTDB test records (mean $\pm$ SD across 3 seeds, 8,848 windows): unlike the beat-based benchmark, both CC and SDR improve monotonically from base2 to base16. (d) Beat-based nested noise-pool restriction (3 seeds; error bars: 95\% paired-$t$ CIs; dotted line: $-0.005$ margin): shrinking the 634-window training pool intensifies the base16 $-$ base6 penalty monotonically on BW and MA while remaining strongly negative across all fractions on EM.}
\label{fig:ecg_cap}
\end{figure}

Table~\ref{tab:capacity} summarizes the primary capacity contrasts (base6$\to$8 and base8$\to$16). On beat-based BW, base6$\to$8 ($+0.0005$, 95\% CI $[-0.0018, +0.0028]$) and base8$\to$16 ($-0.0032$, $[-0.0048, -0.0016]$) fall inside $\pm 0.005$ CC, although base8$\to$16 is strictly negative and crosses into a detected decrease under the Bonferroni-adjusted 99.17\% CI. On MA and EM, scaling from base8 to base16 causes significant decreases in both CC and SDR: MA drops by $-0.0098$ CC ($[-0.0128, -0.0069]$, SDR $-0.37$~dB) and EM drops by $-0.0379$ CC ($[-0.0475, -0.0282]$, SDR $-1.35$~dB).

\begin{table}[tbp]
\centering
\caption{Primary capacity contrasts (base6$\to$8 and base8$\to$16, 5 matched seeds) for beat-based ECG (nv1) and sEMG. Conservative equivalence margins: $\pm 0.005$ CC, $\pm 0.10$~dB SDR. Source column reports the 20,000-resample two-way pigeonhole bootstrap 95\% CI (record/subject $\times$ noise window) for CC. Four-way classification keeps meaningful decreases distinct from practical equivalence.}
\label{tab:capacity}
\scriptsize
\setlength{\tabcolsep}{3.0pt}
\begin{tabular}{@{}lllrrrrl@{}}
\toprule
\textbf{Task} & \textbf{Contr.} & \textbf{Met.} & \textbf{Inc.} & \textbf{Seed 95\% CI} & \textbf{$m^*$} & \textbf{Two-way source CI} & \textbf{Classification} \\
\midrule
ECG BW & b6$\to$8 & CC & $+0.0005$ & $[-0.0018, +0.0028]$ & 0.0028 & $[-0.0006, +0.0016]$ & Equivalence \\
ECG BW & b8$\to$16 & CC & $-0.0032$ & $[-0.0048, -0.0016]$ & 0.0048 & $[-0.0070, +0.0006]$ & Equivalence$^\dagger$ \\
ECG EM & b6$\to$8 & CC & $-0.0029$ & $[-0.0118, +0.0060]$ & 0.0118 & $[-0.0067, +0.0013]$ & Inconclusive \\
ECG EM & b8$\to$16 & CC & $-0.0379$ & $[-0.0475, -0.0282]$ & 0.0475 & $[-0.0485, -0.0266]$ & Decrease \\
ECG MA & b6$\to$8 & CC & $-0.0010$ & $[-0.0040, +0.0020]$ & 0.0040 & $[-0.0024, +0.0008]$ & Equivalence \\
ECG MA & b8$\to$16 & CC & $-0.0098$ & $[-0.0128, -0.0069]$ & 0.0128 & $[-0.0150, -0.0045]$ & Decrease \\
\midrule
sEMG ECG & b6$\to$8 & CC & $-0.0005$ & $[-0.0047, +0.0037]$ & 0.0047 & $[-0.0014, +0.0001]$ & Equivalence \\
sEMG ECG & b8$\to$16 & CC & $+0.0022$ & $[-0.0002, +0.0047]$ & 0.0047 & $[+0.0008, +0.0040]$ & Equivalence$^\ddagger$ \\
sEMG ECG & b8$\to$16 & SDR & $+0.45$ & $[+0.12, +0.78]$ & 0.78 & -- & Positive gain$^*$ \\
sEMG MOA & b6$\to$8 & CC & $+0.0011$ & $[-0.0000, +0.0022]$ & 0.0022 & $[+0.0009, +0.0013]$ & Equivalence \\
sEMG MOA & b8$\to$16 & CC & $+0.0006$ & $[+0.0000, +0.0011]$ & 0.0011 & $[+0.0004, +0.0008]$ & Equivalence \\
sEMG MOA & b8$\to$16 & SDR & $+0.20$ & $[+0.04, +0.36]$ & 0.36 & -- & Positive gain \\
sEMG PLI & b6$\to$8 & CC & $+0.0023$ & $[-0.0001, +0.0047]$ & 0.0047 & $[+0.0019, +0.0027]$ & Equivalence \\
sEMG PLI & b8$\to$16 & CC & $+0.0035$ & $[+0.0022, +0.0047]$ & 0.0047 & $[+0.0030, +0.0040]$ & Equivalence \\
sEMG PLI & b8$\to$16 & SDR & $+1.14$ & $[+0.79, +1.50]$ & 1.50 & -- & Positive gain \\
\bottomrule
\end{tabular}
\vspace{2pt}
\parbox{\textwidth}{\footnotesize $^\dagger$Strictly negative; becomes Decrease under Bonferroni 99.17\% CI and Inconclusive under two-way source bootstrap. $^\ddagger$Seed-plus-source interval $[-0.0011, +0.0056]$ (Inconclusive). $^*$Vanishes to $-0.006$~dB at 75 epochs (Section~\ref{sec:emg_cap}).}
\end{table}

Three complementary analyses support finite-noise-pool overfitting as an explanation for the beat-based ECG decline:
\begin{enumerate}
    \item \textbf{Nested noise-pool restriction (Figure~\ref{fig:ecg_cap}d).} Cutting the training noise pool from 100\% (634 windows) to 50\% (317) and 25\% (158) intensifies the base16 $-$ base6 CC penalty monotonically on BW ($-0.0039 \to -0.0163 \to -0.0665$) and MA ($-0.0096 \to -0.0293 \to -0.0404$), while EM remains strongly negative at all fractions ($-0.0434, -0.0549, -0.0537$). Contrast classifications also shift between nv1 and nv2 across seven 3-seed contrasts.
    \item \textbf{Optimization sensitivity (2 seeds).} Tripling training from 25 to 75 epochs worsens base16 test SDR on every task and seed (mean $\Delta\text{SDR}$: $-0.71$~dB BW, $-1.73$~dB EM, $-1.48$~dB MA) while improving base4 on BW and MA and preserving negative base16 $-$ base6 gaps in all runs. Lowering the learning rate to $3\times 10^{-4}$ partially reduces base16 degradation on EM (+1.17~dB) and MA (+0.24~dB) without surpassing base5--base6.
    \item \textbf{Source-cluster uncertainty.} Two-way resampling over the 14 QTDB test records and 634 NSTDB test windows widens intervals substantially (flipping BW base8$\to$16 to inconclusive, $[-0.0070, +0.0006]$), whereas EM and MA base8$\to$16 remain decreases under every resampling scheme.
\end{enumerate}

In sharp contrast, the continuous-window mixed-noise ECG benchmark exhibits monotonic improvement with capacity across all five tested widths (Figure~\ref{fig:ecg_cap}c; full metrics in Appendix~\ref{app:capacity}, Table~\ref{tab:app_ecg_cont}): on the 8,848 held-out QTDB test windows (3 seeds), test CC rises steadily from 0.797 $\pm$ 0.028 (base2) through 0.855 $\pm$ 0.007 (base4), 0.864 $\pm$ 0.005 (base6), and 0.870 $\pm$ 0.004 (base8) to 0.881 $\pm$ 0.001 (base16), while SDR increases from 5.24 to 8.09~dB. Thus, the 3.2K--6.5K capacity ceiling is specific to the single-artifact beat-based QTDB+NSTDB protocol rather than a universal property of ECG denoising.

\subsection{sEMG Capacity Curves, Metric Divergence, and Optimization Sensitivity}
\label{sec:emg_cap}

sEMG follows a distinct capacity trajectory (Figure~\ref{fig:emg_cap}a--b; full values in Appendix~\ref{app:capacity}, Table~\ref{tab:app_emg_fine}). Waveform CC rises rapidly from base1 (0.837--0.874) to base4 (0.963--0.978) and shows diminishing returns beyond approximately 3.2K parameters: all six tested primary adjacent CC contrasts (base6$\to$8 and base8$\to$16 across ecg, moa, pli) satisfy formal practical equivalence within $\pm 0.005$ (Table~\ref{tab:capacity}), as do extended adjacent contrasts base16$\to$24, base24$\to$32, and base16$\to$32 (3 seeds, all $|\Delta\text{CC}| \le 0.0030$). However, adjacent equivalence does not imply global equivalence across the entire ladder: PLI CC edges modestly upward cumulatively from 0.978 (base4) to 0.990 (base32), a $+0.012$ multi-step rise exceeding the single-step $\pm 0.005$ margin.

\begin{figure}[tbp]
\centering
\includegraphics[width=\textwidth]{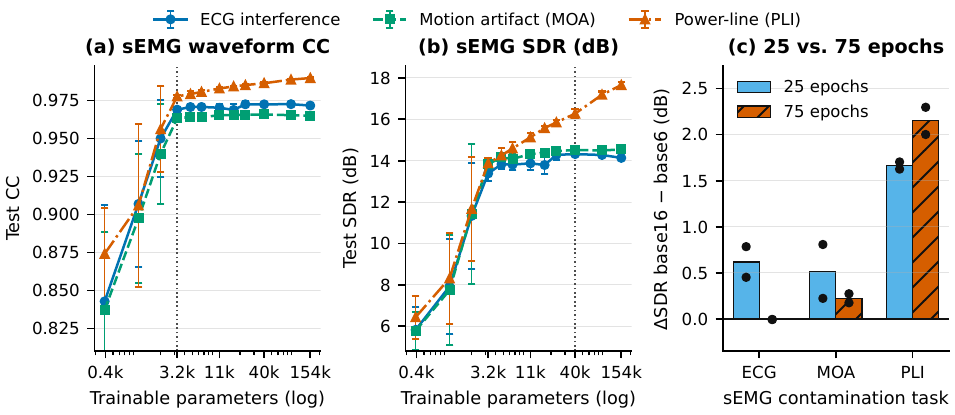}
\caption{Controlled sEMG capacity curves (392--153,718 parameters, base1--base32; mean $\pm$ SD across 5 seeds, or 3 seeds for base10/12/24/32) and training-budget sensitivity on NinaPro DB2. (a) Waveform CC shows diminishing returns beyond base4 (3.2K, dotted line), with formal practical equivalence ($\pm 0.005$) across tested adjacent contrasts from base6 upward while PLI CC edges modestly upward cumulatively through base32. (b) SDR diverges from CC, continuing to rise for PLI across the full ladder to base32 (153.7K). (c) Exploratory 75-epoch sensitivity (base16 $-$ base6 SDR, 2 seeds shown as dots): the ECG-interference advantage vanishes ($+0.62 \to -0.006$~dB), whereas PLI gains intensify ($+1.66 \to +2.15$~dB).}
\label{fig:emg_cap}
\end{figure}

In contrast, SDR and mean-frequency error (MF-RMSE) diverge markedly from waveform CC. From base8 to base16, SDR increases significantly on every task (Table~\ref{tab:capacity}: $+0.45$~dB ecg, $+0.20$~dB moa, $+1.14$~dB pli) alongside MF-RMSE reductions (ecg 8.40 to 6.95~Hz; moa 8.34 to 8.17~Hz; pli 3.15 to 2.66~Hz). On the matched 3-seed subset (seeds 42--44, where base16 PLI SDR is 16.41~dB vs.\ 16.28~dB across 5 seeds), SDR shows little further improvement beyond base16 for ecg ($-0.22$~dB to base32) and moa ($+0.02$~dB), whereas narrowband PLI continues gaining $+0.78$~dB from base16 to base24 (95\% CI $[+0.69, +0.87]$) and $+1.25$~dB to base32 (SDR 17.66~dB, MF-RMSE 2.24~Hz).

The extended-training sensitivity analysis (Figure~\ref{fig:emg_cap}c, 2 seeds; Appendix~\ref{app:controls}) reveals an optimization confound at 25 epochs: extending training to 75 epochs eliminates the base16 $-$ base6 SDR advantage for cardiac interference (from $+0.78 / +0.45$~dB to $-0.006 / -0.007$~dB; $|\Delta\text{CC}| \le 0.0004$), reduces it for MOA ($+0.81 / +0.22 \to +0.17 / +0.27$~dB), and strengthens it for PLI ($+1.63 / +1.70 \to +2.30 / +2.00$~dB). Stratified by test SNR (Appendix~\ref{app:controls}, Table~\ref{tab:app_emg_snr}), base6$\to$16 SDR gains grow with input SNR for ecg ($+0.17$ to $+1.21$~dB) and moa ($+0.21$ to $+0.88$~dB) and remain substantial across all SNRs for PLI ($+1.47$ to $+2.02$~dB).

\subsection{Controlled Baselines, High-Capacity References, and Compute Footprint}

\begin{figure}[tbp]
\centering
\includegraphics[width=\textwidth]{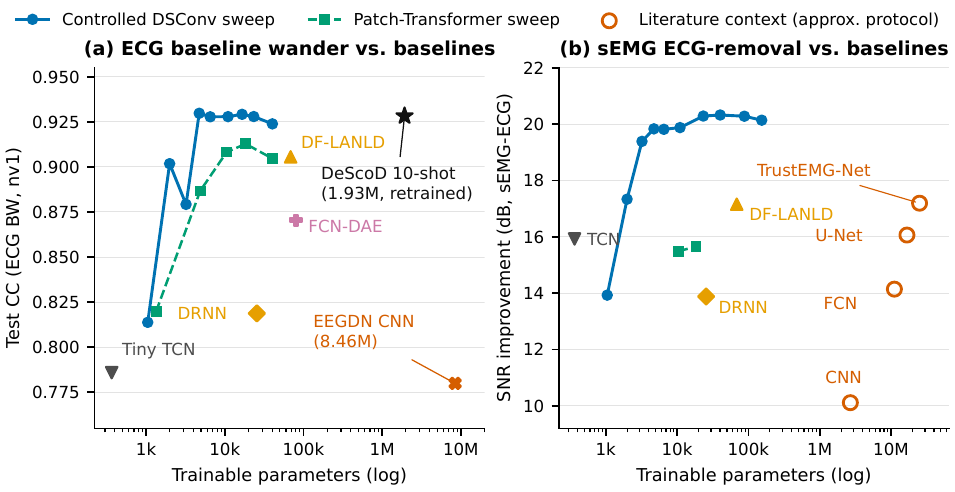}
\caption{Parameter-efficiency trade-offs across controlled baselines and high-capacity references (3 seeds, nv1). (a) ECG baseline wander: controlled DSConv base5--base6 (4.7K--6.5K) outperform all same-pipeline baselines and achieve similar mean CC to 3-seed retrained 10-shot DeScoD-ECG (1.93M; 0.928 vs.\ 0.928); Patch-Transformer exhibits a similar peak-and-drop pattern ($d_{32} \to d_{48}$). (b) sEMG ECG-interference removal: DSConv base4--base16 outperform all same-pipeline baselines; open circles show published sEMG architectures from \cite{wang2025trustemgnet} as approximately protocol-matched literature context (3,180 test segments, 4-fold CV, single and combined mixtures) rather than directly matched comparisons.}
\label{fig:baselines}
\end{figure}

Figure~\ref{fig:baselines} and Table~\ref{tab:baselines} compare the controlled DSConv sweep against all same-pipeline baselines and high-capacity references.

\begin{table}[tbp]
\centering
\caption{Controlled baseline comparison (3 seeds, shared 25-epoch recipe), published-recipe DeScoD-ECG references, and single-thread CPU profiling on an idle AMD EPYC 7313 host (batch size 1; 512 samples ECG, 2,000 samples sEMG). Published sEMG rows are approximately protocol-matched literature context from \cite{wang2025trustemgnet}.}
\label{tab:baselines}
\scriptsize
\setlength{\tabcolsep}{3.0pt}
\resizebox{\textwidth}{!}{%
\begin{tabular}{@{}lrrrrrrr@{}}
\toprule
\textbf{Model (ECG, nv1)} & \textbf{Params} & \textbf{BW CC} & \textbf{BW SSD} & \textbf{EM CC} & \textbf{MA CC} & \textbf{Size (KB)} & \textbf{CPU (ms)} \\
\midrule
Learned FIR (33-tap) & 34 & 0.380 & 23.83 & 0.280 & 0.317 & -- & -- \\
Tiny TCN & 365 & 0.786 & 9.85 & 0.714 & 0.718 & -- & -- \\
DSConv base4 & 3,218 & 0.879 & 6.70 & 0.815 & 0.923 & 33.3 & 1.59 \\
DSConv base5 & 4,732 & \textbf{0.930} & 4.24 & 0.817 & \textbf{0.927} & -- & -- \\
DSConv base6 & 6,532 & 0.928 & 4.37 & 0.814 & 0.927 & 45.8 & 1.67 \\
DSConv base16 & 40,262 & 0.924 & 4.86 & 0.771 & 0.917 & 179.8 & 2.20 \\
Patch-Transformer $d_{16}$ & 4,865 & 0.887 & 7.30 & \textbf{0.821} & 0.910 & -- & -- \\
Patch-Transformer $d_{32}$ & 18,433 & 0.913 & 5.46 & 0.812 & 0.910 & -- & -- \\
Patch-Transformer $d_{48}$ & 40,705 & 0.905 & 6.31 & 0.811 & 0.910 & -- & -- \\
DRNN (port) & 25,537 & 0.819 & 8.45 & 0.799 & 0.854 & -- & -- \\
DeepFilter LANLD (port) & 68,719 & 0.906 & 5.40 & 0.773 & 0.908 & -- & -- \\
FCN-DAE (port) & 80,223 & 0.870 & 7.56 & 0.720 & 0.869 & -- & -- \\
EEGDN Complex CNN & 8,455,424 & 0.780 & 37.70 & 0.665 & 0.818 & -- & -- \\
DeScoD-ECG 1-shot (3-seed retr.) & 1,926,321 & 0.904 & 5.40 & -- & -- & -- & -- \\
DeScoD-ECG 10-shot (3-seed retr.) & 1,926,321 & 0.928 & \textbf{4.05} & -- & -- & -- & -- \\
DeScoD-ECG 10-shot (released wts) & 1,926,321 & 0.929 & 4.04 & -- & -- & -- & -- \\
\midrule
\textbf{Model (sEMG, 2,000 samples)} & \textbf{Params} & \textbf{ECG $\Delta$SNR} & \textbf{ECG MF} & \textbf{MOA $\Delta$SNR} & \textbf{PLI $\Delta$SNR} & \textbf{Size (KB)} & \textbf{CPU (ms)} \\
\midrule
Learned FIR (33-tap) & 34 & 13.43 & 17.15 & 17.90 & 12.16 & -- & -- \\
Tiny TCN & 365 & 15.92 & 12.01 & 16.32 & 13.58 & -- & -- \\
DSConv base4 & 3,218 & 19.21 & 8.64 & 19.84 & 20.03 & 33.3 & 1.91 \\
DSConv base6 & 6,532 & 19.67 & 8.62 & 20.00 & 20.53 & 45.8 & 2.20 \\
DSConv base16 & 40,262 & \textbf{20.35} & \textbf{6.91} & \textbf{20.53} & 22.41 & 179.8 & 3.78 \\
DSConv base24 & 87,838 & 20.28 & 6.44 & 20.50 & \textbf{23.19} & 366.8 & 5.15 \\
Patch-Transformer $d_{32}$ & 18,433 & 15.65 & 13.60 & 15.86 & 12.45 & -- & -- \\
DRNN (port) & 25,537 & 13.88 & 12.86 & 14.72 & 13.66 & -- & -- \\
DeepFilter LANLD (port) & 68,719 & 17.15 & 10.37 & 18.47 & 17.22 & -- & -- \\
FCN (lit.\ context \cite{wang2025trustemgnet}) & 11,150,000 & 14.14 & 10.73 & 13.43 & 15.84 & -- & -- \\
U-Net (lit.\ context \cite{wang2025trustemgnet}) & 16,720,000 & 16.06 & 14.50 & 14.73 & 20.86 & -- & -- \\
TrustEMG-Net (lit.\ context \cite{wang2025trustemgnet}) & 25,120,000 & 17.19 & 11.67 & 15.89 & 21.45 & -- & -- \\
\bottomrule
\end{tabular}%
}
\end{table}

On beat-based ECG BW and MA, no controlled baseline exceeds DSConv base5--base6. Importantly, the Patch-Transformer sweep exhibits a similar non-monotonic peak-and-drop pattern across both investigated architecture families: on BW it peaks at $d_{32}$ (18.4K, CC 0.913) and drops at $d_{48}$ (40.7K, CC 0.905); on EM, compact Patch-Transformers ($d_{16}$: 4.87K, CC 0.821, SDR 5.53~dB; $d_{24}$: 10.56K, CC 0.819) slightly outperform DSConv base6 (CC 0.814, SDR 5.24~dB) before declining at $d_{32}$--$d_{48}$ (CC 0.812--0.811). The 8.46M-parameter EEGDN CNN achieves lower reconstruction performance across all tasks (BW CC 0.780, EM 0.665, MA 0.818), and recipe-sensitivity controls (learning rate $3\times 10^{-4}$ or 75 epochs) remain 0.096--0.139 CC below base6 (Appendix~\ref{app:controls}, Table~\ref{tab:app_controls}). Against published-recipe DeScoD-ECG on nv1 BW, base6 (6.5K) exceeds 3-seed retrained 1-shot diffusion (SSD 4.37 vs.\ 5.40~a.u.) and sits 0.32~a.u.\ behind 3-seed retrained 10-shot diffusion (4.05~a.u.; released-weights evaluation 4.04~a.u.; published paper value 3.77~a.u.\ \cite{li2024descod}) at $295\times$ fewer parameters. On pooled nv1+nv2 BW, base6 achieves SSD 4.24~a.u.\ and PRD 40.78\%, improving upon DeepFilter's published LANLD figures (SSD 5.20~a.u., PRD 50.45\%).

On sEMG, no controlled baseline reaches base6 on any task: DeepFilter LANLD (68.7K) trails base6 by 1.5--3.3~dB across tasks, while even a 34-parameter learned FIR filter reaches $\text{SNR}_{\text{imp}}$ 17.9~dB (CC 0.951) on MOA. On an idle AMD EPYC 7313 CPU (single thread, batch size 1), base4--base6 require 33.3--45.8~KB and execute in 1.59--1.67~ms per 512-sample ECG segment (0.64M--1.31M MACs) and 1.91--2.20~ms per 2,000-sample sEMG segment (2.49M--5.10M MACs); exported ONNX checkpoints for base4, base6, and base16 matched PyTorch outputs within $10^{-5}$.

\subsection{Downstream Utility on Unmodified Clinical ECG}

\begin{figure}[tbp]
\centering
\includegraphics[width=\textwidth]{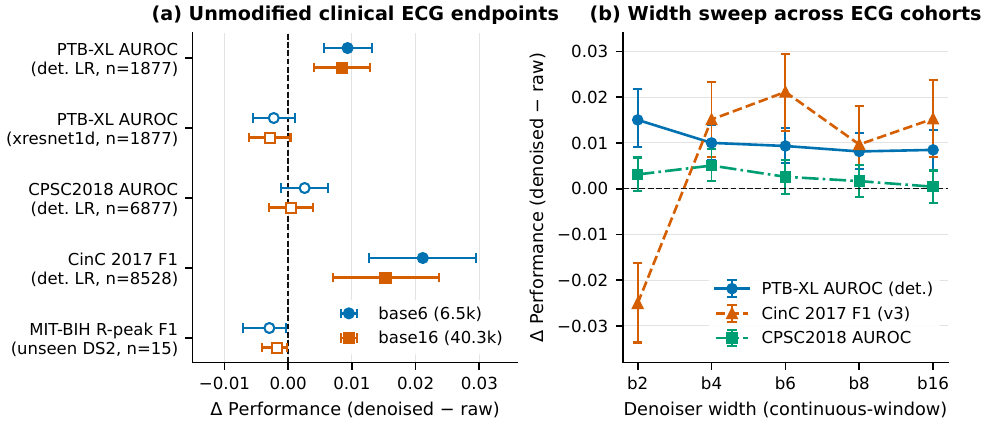}
\caption{Downstream utility on unmodified clinical ECG recordings (\texttt{denoised/denoised} minus \texttt{raw/raw}, 95\% patient/recording bootstrap CIs). (a) Primary deterministic and neural endpoints at base6 (circles) and base16 (squares); filled markers denote BH-FDR-supported effects ($q < 0.05$). (b) Width sweep across deterministic pipelines on PTB-XL, CinC 2017, and CPSC2018 (error bars: 95\% unit bootstrap CIs).}
\label{fig:ecg_ds}
\end{figure}

Applying continuous-window mixed-noise denoisers to unmodified clinical ECG recordings produced small, cohort- and classifier-dependent changes (Figure~\ref{fig:ecg_ds} and Table~\ref{tab:downstream}).

Within the 14-test primary clinical ECG family (7 endpoints $\times$ base6 and base16), only two deterministic endpoints showed BH-FDR-supported (and Holm-supported) gains: PTB-XL macro AUROC rose from 0.8673 raw by $+0.0093$ (base6, 95\% CI $[+0.0056, +0.0132]$, $q < 0.001$) and $+0.0085$ (base16, $[+0.0041, +0.0128]$, $q < 0.001$), and CinC 2017 challenge F1 rose from 0.6763 raw by $+0.0211$ (base6, $[+0.0126, +0.0295]$, $q < 0.001$) and $+0.0153$ (base16, $[+0.0070, +0.0237]$, $q = 0.001$; consistent on original labels). On the post-hoc provenance-restricted MIT-BIH DS2 subset ($n=15$ strictly unseen-parent records; Table~\ref{tab:downstream}), no endpoint showed statistically supported changes after BH-FDR ($q \ge 0.079$: R-peak F1 raw 0.9979, base6 $-0.0029$, base16 $-0.0018$, $q=0.264$; beat accuracy raw 0.8338, base6 $+0.0273$, $q=0.079$, base16 $+0.0038$, $q=0.864$; HRV SDNN/RMSSD error all $q \ge 0.266$). Results were consistent after excluding record 202 ($n=14$, all $q \ge 0.088$) and on the full 22-record standard DS2 benchmark (Table~\ref{tab:downstream}, all $q \ge 0.21$), as well as for QTDB 4-fiducial delineation MAE (raw 34.5~ms; 3-fiducial sensitivity raw 24.0~ms, $q \ge 0.33$).

Two external controls demonstrate that the modest deterministic ECG gains do not generalize universally:
\begin{itemize}
    \item \textbf{Classifier dependence (PTB-XL \texttt{xresnet1d101}).} With the end-to-end residual classifier (raw macro AUROC 0.9313), denoising showed no statistically supported improvement and trended slightly negative: $-0.0023$ (base6, 95\% CI $[-0.0056, +0.0011]$, $q=0.19$) and $-0.0029$ (base16, $[-0.0062, +0.0004]$, $q=0.16$).
    \item \textbf{External cohort replication (CPSC2018) and low-capacity degradation.} Replicating the deterministic 12-lead pipeline on CPSC2018 (raw AUROC 0.9057, $n=6,877$) yielded near-null changes at base6 ($+0.0026$, $q=0.35$) and base16 ($+0.0004$, $q=0.81$), with only exploratory base4 reaching $+0.0050$ ($q=0.010$). Conversely, under-parameterized base2 degraded CinC 2017 F1 ($-0.0251$, $q < 0.001$) and raised the predicted Noisy share ($+0.0168$, $q < 0.001$).
\end{itemize}

\begin{table}[tbp]
\centering
\caption{Summary of matched downstream utility (\texttt{denoised/denoised} minus \texttt{reference/reference} at base6 and base16) across unmodified clinical ECG cohorts, NinaPro DB2 sEMG, and real GRABMyo sEMG. Bold denotes BH-FDR $q < 0.05$ within each declared family.}
\label{tab:downstream}
\scriptsize
\setlength{\tabcolsep}{2.8pt}
\resizebox{\textwidth}{!}{%
\begin{tabular}{@{}lllrlrlr@{}}
\toprule
\textbf{Cohort / Endpoint} & \textbf{Pipeline} & \textbf{Unit ($n$)} & \textbf{Ref.} & \textbf{$\Delta$ base6 [95\% CI]} & \textbf{$q_{\text{BH}}$} & \textbf{$\Delta$ base16 [95\% CI]} & \textbf{$q_{\text{BH}}$} \\
\midrule
PTB-XL AUROC & Det.\ LR & Pat.\ (1,877) & 0.867 & $\mathbf{+0.0093}$ $[+0.0056,+0.0132]$ & $<0.001$ & $\mathbf{+0.0085}$ $[+0.0041,+0.0128]$ & $<0.001$ \\
PTB-XL AUROC & \texttt{xresnet1d101} & Pat.\ (1,877) & 0.931 & $-0.0023$ $[-0.0056,+0.0011]$ & 0.190 & $-0.0029$ $[-0.0062,+0.0004]$ & 0.159 \\
CPSC2018 AUROC & Det.\ LR & Rec.\ (6,877) & 0.906 & $+0.0026$ $[-0.0012,+0.0063]$ & 0.352 & $+0.0004$ $[-0.0031,+0.0040]$ & 0.815 \\
CinC 2017 F1 & Det.\ LR & Rec.\ (8,528) & 0.676 & $\mathbf{+0.0211}$ $[+0.0126,+0.0295]$ & $<0.001$ & $\mathbf{+0.0153}$ $[+0.0070,+0.0237]$ & 0.001 \\
MIT-BIH R-peak F1 & Pan--Tomp. & Rec.\ (15)$^*$ & 0.998 & $-0.0029$ $[-0.0071,-0.0004]$ & 0.264 & $-0.0018$ $[-0.0041,-0.0002]$ & 0.264 \\
MIT-BIH beat acc. & Morph.+LDA & Rec.\ (15)$^*$ & 0.834 & $+0.0273$ $[+0.0072,+0.0492]$ & 0.079 & $+0.0038$ $[-0.0397,+0.0410]$ & 0.864 \\
MIT-BIH R-peak F1 (full) & Pan--Tomp. & Rec.\ (22) & 0.998 & $-0.0029$ $[-0.0063,-0.0004]$ & 0.210 & $-0.0023$ $[-0.0051,-0.0002]$ & 0.227 \\
MIT-BIH beat acc.\ (full) & Morph.+LDA & Rec.\ (22) & 0.715 & $+0.0342$ $[-0.0201,+0.0872]$ & 0.332 & $-0.0228$ $[-0.1029,+0.0357]$ & 0.578 \\
QTDB 4-fid.\ MAE & NK2 DWT & Rec.\ (14) & 34.5 & $-3.18$ $[-8.85,+1.84]$ & 0.332 & $-1.58$ $[-7.08,+3.20]$ & 0.578 \\
\midrule
DB2 ($-10$~dB) & TD-LDA & Subj.\ (40) & 0.578 & $\mathbf{+0.0691}$ $[+0.0622,+0.0758]$ & $<0.001$ & $\mathbf{+0.0854}$ $[+0.0775,+0.0931]$ & $<0.001$ \\
DB2 ($0$~dB) & TD-LDA & Subj.\ (40) & 0.648 & $\mathbf{+0.0305}$ $[+0.0249,+0.0359]$ & $<0.001$ & $\mathbf{+0.0430}$ $[+0.0371,+0.0492]$ & $<0.001$ \\
DB2 ($+10$~dB) & TD-LDA & Subj.\ (40) & 0.694 & $\mathbf{+0.0046}$ $[+0.0019,+0.0072]$ & 0.002 & $\mathbf{+0.0059}$ $[+0.0035,+0.0084]$ & $<0.001$ \\
DB2 (clean) & TD-LDA & Subj.\ (40) & 0.718 & $-0.0006$ $[-0.0017,+0.0005]$ & 0.303 & $-0.0006$ $[-0.0016,+0.0004]$ & 0.303 \\
DB2 ($-10$~dB) & TD-SVM & Subj.\ (40) & 0.536 & $\mathbf{+0.1234}$ $[+0.1160,+0.1309]$ & $<0.001$ & $\mathbf{+0.1466}$ $[+0.1390,+0.1544]$ & $<0.001$ \\
DB2 ($-10$~dB) & 1D-CNN & Subj.\ (40) & 0.531 & $\mathbf{+0.1480}$ $[+0.1405,+0.1553]$ & $<0.001$ & $\mathbf{+0.1624}$ $[+0.1546,+0.1702]$ & $<0.001$ \\
DB2 (clean) & 1D-CNN & Subj.\ (40) & 0.743 & $+0.0009$ $[-0.0004,+0.0022]$ & 0.210 & $+0.0014$ $[+0.0000,+0.0031]$ & 0.095 \\
\midrule
GRABMyo fore.\ vs.\ raw & TD-LDA & Subj.\ (43) & 0.773 & $+0.0018$ $[-0.0015,+0.0053]$ & 0.412 & $+0.0030$ $[-0.0005,+0.0072]$ & 0.356 \\
GRABMyo fore.\ vs.\ conv & TD-LDA & Subj.\ (43) & 0.769 & $\mathbf{+0.0059}$ $[+0.0028,+0.0092]$ & 0.004 & $\mathbf{+0.0071}$ $[+0.0037,+0.0108]$ & 0.003 \\
GRABMyo wrist vs.\ raw & TD-LDA & Subj.\ (43) & 0.791 & $-0.0018$ $[-0.0050,+0.0018]$ & 0.412 & $-0.0024$ $[-0.0061,+0.0014]$ & 0.403 \\
GRABMyo wrist vs.\ conv & TD-LDA & Subj.\ (43) & 0.788 & $+0.0010$ $[-0.0021,+0.0041]$ & 0.627 & $+0.0003$ $[-0.0032,+0.0038]$ & 0.852 \\
\bottomrule
\end{tabular}%
}
\vspace{2pt}
\parbox{\textwidth}{\footnotesize $^*$Post-hoc provenance-restricted MIT-BIH DS2 subset ($n=15$ strictly unseen-parent records; excluding record 202 gives $n=14$ with consistent non-significant results, $q \ge 0.088$). Full 22-record standard DS2 benchmark rows are shown directly below.}
\end{table}

\subsection{Downstream Utility on sEMG Gesture Recognition}

\begin{figure}[tbp]
\centering
\includegraphics[width=\textwidth]{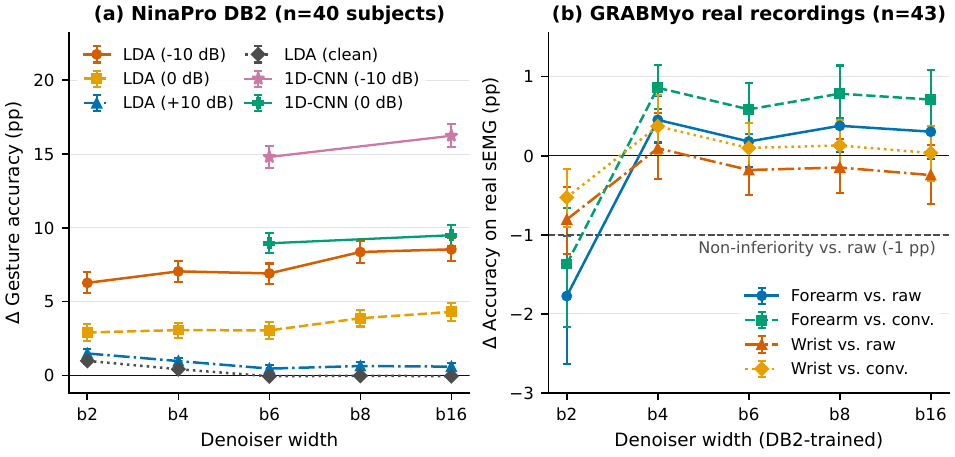}
\caption{Within-subject sEMG gesture-recognition utility across denoiser widths (95\% subject bootstrap CIs). (a) NinaPro DB2 ($n=40$): large accuracy gains at $-10$ and $0$~dB for both TD-LDA and 1D-CNN, alongside slight clean-recording gains exclusively at narrow widths (base2/base4). (b) GRABMyo real recordings ($n=43$ subjects unseen during denoiser training, evaluated under within-subject TD-LDA): base4--base16 satisfy the $-1$~pp non-inferiority margin relative to raw recordings (dashed line) and exceed conventional band-pass plus notch filtering on the forearm, whereas base2 degrades accuracy.}
\label{fig:emg_ds}
\end{figure}

On NinaPro DB2 ($n=40$ subjects; Figure~\ref{fig:emg_ds}a and Table~\ref{tab:downstream}), matched denoising delivered large accuracy gains under heavy and moderate synthetic contamination across linear and neural decoders. At $-10$~dB, TD-LDA gained $+6.91$~pp at base6 and $+8.54$~pp at base16 from 0.578 raw; TD-SVM gained $+12.34$ and $+14.66$~pp; and the end-to-end 1D-CNN gained $+14.80$ and $+16.24$~pp from 0.531 raw (all $q < 0.001$). In post-hoc paired comparisons on DB2, base16 exceeded base6 by $+1.63$~pp (TD-LDA) and $+2.33$~pp (TD-SVM) at $-10$~dB, and by $+1.26$ and $+2.41$~pp at $0$~dB (all $q < 0.001$). On unmodified DB2 recordings, base6--base16 showed no statistically supported changes across the three decoders ($|\Delta| \le 0.14$~pp, $q \ge 0.095$), whereas narrow widths slightly improved clean accuracy (base2: $+0.98$~pp TD-LDA, $+0.99$~pp TD-SVM; base4: $+0.42$ and $+0.50$~pp, all $q < 0.001$) and gave the largest gains at $+10$~dB (base2: $+1.48$~pp vs.\ base6: $+0.46$~pp).

Transferred zero-shot to 43 unseen GRABMyo subjects with real recordings (Figure~\ref{fig:emg_ds}b), base4--base16 satisfied the $-1$~pp non-inferiority criterion relative to raw signals: lower 95\% CI bounds for \texttt{denoised} minus \texttt{raw} at base6 and base16 were $-0.15$ and $-0.05$~pp on the forearm and $-0.50$ and $-0.61$~pp on the wrist (all $>-1$~pp). On the forearm, neural denoising outperformed conventional 20--450~Hz band-pass plus notch filtering by $+0.59$~pp (base6, $q=0.004$) and $+0.71$~pp (base16, $q=0.003$; conventional filtering reduced raw forearm accuracy by $-0.41$~pp, $q=0.001$). Exploratory base16 $-$ base6 differences on GRABMyo were small and non-significant (forearm $+0.12$~pp, $p=0.079$; wrist $-0.06$~pp, $p=0.50$). Cross-day sessions showed the same non-inferior pattern, whereas base2 degraded real-recording accuracy (forearm within-session $-1.77$~pp, cross-day $-2.21$~pp; wrist within-session $-0.81$~pp, all $q \le 0.003$). No-reference spectral indices indicated that base4--base16 substantially reduced low-frequency ($<20$~Hz) signal power relative to raw recordings (forearm median fraction $0.0175 \to 0.0043$ at base6, $-0.61 \log_{10}$; wrist $0.0322 \to 0.0061$, $-0.72 \log_{10}$), while leaving the pre-notched 60~Hz band essentially unchanged ($\pm 0.01 \log_{10}$).

\subsection{Same-Condition Metric--Utility Associations and the sEMG SNR Sign Reversal}

On synthetically contaminated clinical ECG records (independent unseen-parent MIT-BIH DS2 R-peak F1, $n=15$, and QTDB delineation MAE, $n=14$; 5 metrics $\times$ 6 conditions $= 30$ tests), 20 of 30 within-record CR1 slopes were positive and BH-FDR-supported (25 of 30 under Webb wild cluster bootstrap and two-stage Wilcoxon; Figure~\ref{fig:mu}a; full slopes, 95\% CIs, and five-method $q$-values in Appendix~\ref{app:mu}, Table~\ref{tab:app_ecg_mu}). Associations were strongest under electrode motion (e.g., MIT-BIH CC slope $+0.0208$ F1 per within-SD, 95\% CI $[+0.0157, +0.0260]$, 15/15 records positive; QTDB SDR $+3.63$~ms reduction per within-SD, $[+2.46, +4.80]$, 14/14 positive). Under muscle artifact on the 15 unseen-parent MIT-BIH records, point slopes remained positive in 13--14/15 records ($+0.0060$ to $+0.0076$, borderline under CR1 at $q=0.053$--$0.065$, BH-FDR-supported under Webb wild cluster bootstrap at $q \le 0.038$ and two-stage Wilcoxon at $q \le 0.050$; on the full 22-record DS2 set, 25 of 30 CR1 and 21 of 30 CR2 slopes are supported). Under baseline wander on MIT-BIH, associations were absent across all five metrics ($q \ge 0.35$), potentially reflecting the low-frequency attenuation of Pan--Tompkins preprocessing. Under CR2/Satterthwaite small-cluster inference ($n=14$--$15$), 12 of 30 independent-subset slopes remained BH-FDR-supported (Table~\ref{tab:app_ecg_mu}).

\begin{figure}[tbp]
\centering
\includegraphics[width=\textwidth]{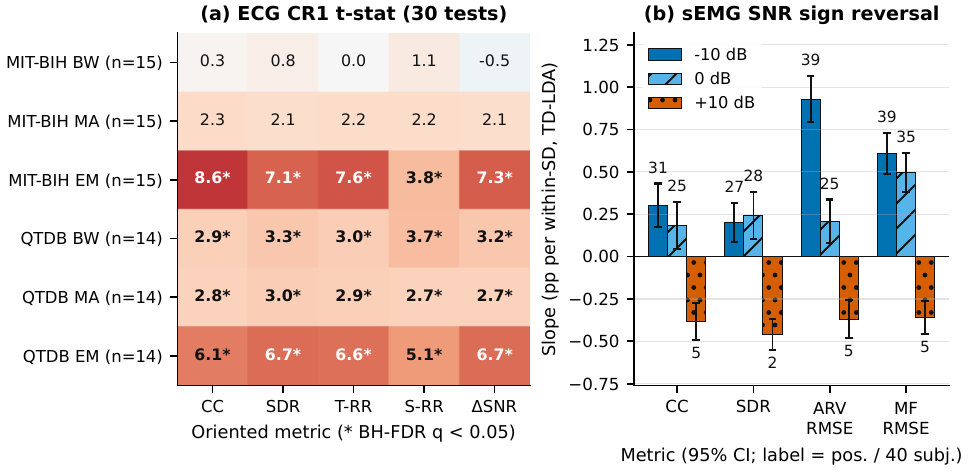}
\caption{Same-condition metric--utility associations. (a) All 30 ECG within-record CR1 $t$-statistics across five oriented metrics (CC, SDR, T-RRMSE, S-RRMSE, $\Delta$SNR) on the provenance-restricted independent MIT-BIH DS2 subset ($n=15$) and QTDB delineation MAE ($n=14$); asterisks mark BH-FDR $q < 0.05$ (MIT-BIH MA $n=15$ is additionally supported under Webb wild cluster bootstrap and Wilcoxon, Table~\ref{tab:app_ecg_mu}). (b) sEMG within-subject TD-LDA slopes on NinaPro DB2 ($n=40$, pp per within-subject metric SD, with 95\% CR1 CIs and count of positive-slope subjects): all four metrics are positive at $-10$ and $0$~dB and reverse to negative at $+10$~dB.}
\label{fig:mu}
\end{figure}

In sEMG, the same-condition analysis for feature-based linear decoders on NinaPro DB2 ($n=40$ subjects, 600 observations per classifier $\times$ SNR) revealed a consistent SNR-dependent sign reversal (Figure~\ref{fig:mu}b; full 95\% CIs and all five inference methods in Appendix~\ref{app:mu}, Table~\ref{tab:app_emg_mu}). Every one of the 24 tests (2 classifiers $\times$ 3 SNRs $\times$ 4 metrics) satisfied BH-FDR $q < 0.05$ under all five inference methods (CR1, CR2/Satterthwaite, Webb wild cluster bootstrap, two-stage $t$, and two-stage Wilcoxon), with opposite signs across SNR:
\begin{itemize}
    \item \textbf{Low and moderate SNR ($-10$ and $0$~dB):} Higher reconstruction fidelity is associated with higher gesture accuracy, with feature-domain metrics tracking accuracy more steeply than waveform metrics at $-10$~dB (TD-LDA: ARV-RMSE $+0.93$~pp per within-SD, 95\% CI $[+0.79, +1.07]$, 39/40 subjects positive; MF-RMSE $+0.61$~pp, $[+0.49, +0.73]$, 39/40 positive; CC $+0.30$~pp, $[+0.17, +0.43]$; SDR $+0.20$~pp, $[+0.08, +0.32]$; TD-SVM: MF-RMSE $+1.12$~pp, $[+0.99, +1.25]$, 40/40 positive).
    \item \textbf{High SNR ($+10$~dB):} Every slope reverses to negative across both classifiers and all four metrics (TD-LDA: SDR $-0.46$~pp per within-SD, 95\% CI $[-0.55, -0.37]$, 2/40 subjects positive; CC $-0.38$~pp, $[-0.49, -0.28]$, 5/40 positive; ARV-RMSE $-0.37$~pp; MF-RMSE $-0.36$~pp; TD-SVM: SDR $-0.29$~pp, $[-0.40, -0.18]$, 8/40 positive).
\end{itemize}
Note that denoiser width and reconstruction fidelity co-vary across the 15 evaluated checkpoints; we interpret this reversal alongside the clean-recording narrow-width pattern in Section~\ref{sec:disc_utility}.

\section{Discussion}

\subsection{Capacity Behavior Across Biosignal Modalities and Benchmark Protocols}

Two factors may contribute to the distinct capacity regimes observed across our ECG and sEMG experiments (see Appendix~\ref{app:provenance}, Table~\ref{tab:app_crossmod}, for a side-by-side summary with our prior EEG study \cite{bindra2026eeg}).

First, within ECG itself, the sharp contrast between beat-based degradation at base16 and continuous-window monotonic improvement to base16 (Figure~\ref{fig:ecg_cap}a--c) underscores that capacity behavior depends heavily on benchmark construction rather than reflecting a single universal parameter ceiling for ECG denoising. The two protocols differ simultaneously in segmentation (zero-padded single beats vs.\ continuous unpadded windows), contamination diversity (634 windows of one artifact type vs.\ 1,902 distinct windows drawn uniformly across BW, EM, and MA), and validation structure (beat-level random split vs.\ record-level split), so we do not attribute the reversal in capacity slope to any single factor without dedicated factorial ablations. Nevertheless, on the widely used beat-based QTDB+NSTDB split, the similar peak-and-drop curve in both DSConv and Patch-Transformer families, the monotonic escalation of the base16 penalty when shrinking the BW and MA training pools to 25\%, and the worsening of base16 under 75 epochs provide converging evidence consistent with finite-noise-pool overfitting. This highlights an important consideration for recent 2025--2026 NSTDB-based ECG studies \cite{li2025tfcdiff,morel2026drnetmamba,ma2026efibnet}: preventing direct train--test window overlap is necessary but not sufficient when training noise diversity comes from a single 15-minute recording half.

Second, on sEMG, our width ladder and 75-epoch sensitivity clarify a tension noted in recent state-space sEMG work \cite{liu2024msemg,shih2026ssemgnet}, where waveform gains did not automatically yield lower spectral frequency error. Across 392 to 153,718 parameters, CC improvements diminish beyond base4 whereas narrowband PLI rejection continues improving SDR and MF-RMSE across the full ladder to base32 (153.7K). Conversely, for broadband cardiac interference in sEMG, the apparent SDR advantage of base16 over base6 at 25 epochs disappears at 75 epochs, demonstrating that optimization budgets must be checked before attributing benchmark gains to parameter count.

\subsection{Downstream Utility and the Role of Signal Fidelity}
\label{sec:disc_utility}

Our downstream results reinforce recent cautions in clinical ECG benchmarking \cite{morel2026drnetmamba,ma2026efibnet} while highlighting why reconstruction fidelity and classifier architecture interact. On unmodified clinical ECG, handcrafted feature pipelines (PTB-XL logistic regression, CinC 2017) gain modestly from front-end denoising, likely because R-peak, fiducial, and band-power extractors are sensitive to drift and transients. By contrast, an end-to-end residual classifier (\texttt{xresnet1d101}) achieves higher baseline macro AUROC and shows no statistically supported improvement from denoising, suggesting it may be less sensitive to the artifacts addressed by the denoiser. Furthermore, even deterministic pipelines show cohort dependence, with CPSC2018 yielding near-null changes at base6 and base16.

In sEMG, matched denoising provides substantial value under severe synthetic contamination for both linear and 1D-CNN decoders, and transfers zero-shot to real GRABMyo recordings while satisfying the prespecified non-inferiority criterion relative to raw signals and outperforming standard 20--450~Hz band-pass plus notch filtering on the forearm. Conventional filtering and learned denoising present distinct trade-offs: fixed band-pass and notch filters reduced raw GRABMyo forearm accuracy by $-0.41$~pp, whereas learned denoisers reduced low-frequency signal power ($<20$~Hz) while preserving forearm decoding accuracy, though frequency-specific signal distortion was not directly quantified. Yet the $+10$~dB sign reversal in DB2, together with the slight clean-recording accuracy gains exclusively at narrow widths (base2/base4), indicates that clean-reference fidelity is not a universal surrogate for myoelectric control: at high SNR, lower-fidelity, mildly smoothing models can yield better linear-classifier accuracy, potentially through feature stabilization of Hudgins time-domain statistics (WL, ZC, SSC), even while degrading sample-wise fidelity against the unsmoothed target.

\subsection{Practical Engineering Guidance}

Our sweeps identify favorable reconstruction--complexity trade-offs for resource-constrained biosignal systems. For single-lead beat-based ECG denoising, base5--base6 (4.7K--6.5K parameters, 46~KB, 1.67~ms single-thread CPU execution) avoids both low-capacity degradation (base1--base2) and beat-benchmark overfitting at 40K+ parameters without iterative diffusion overhead. For continuous multi-artifact ECG enhancement, base16 (40.3K, 2.20~ms) improves reconstruction, though downstream clinical gains remain modest and pipeline-dependent. For sEMG enhancement, base6 (6.5K, 2.20~ms) provides favorable performance--complexity trade-offs for broadband artifact suppression and real GRABMyo decoding, whereas larger models (base16, 40.3K, 180~KB, 3.78~ms) may be preferable when severe low-SNR recovery or narrowband PLI attenuation is prioritized without dedicated notch filtering. Across both modalities, capacity should be selected via end-to-end downstream validation rather than clean-reference metrics alone.

\subsection{Limitations}

Three groups of limitations bound our findings. First, regarding \emph{data and source provenance}, beat-based ECG reconstruction conclusions apply to the 14 QTDB test records and 634 non-overlapping windows from single 15-minute NSTDB recordings rather than independent multi-session noise corpora, and nested prefix restrictions cannot fully separate pool size from temporal composition; similarly, sEMG MOA test noise uses channel 2 of the NSTDB recording whose channel 1 trains models \cite{wang2025trustemgnet}. QTDB provenance revealed parent-recording overlap with 7 of 22 MIT-BIH DS2 records; prioritizing the 15 unseen DS2 records yielded consistent findings, though retraining continuous denoisers on a MIT-BIH-excluded QTDB split remains desirable. In sEMG, NinaPro DB2 ``clean'' signals are band-passed real recordings retaining intrinsic noise and GRABMyo is pre-notched at 60~Hz, so real-recording evaluations assess pipeline performance rather than latent physiological recovery or real mains interference.

Second, regarding \emph{methodology and baselines}, controlled baselines share the 25-epoch recipe without architecture-specific tuning (except 8.46M CNN controls and published-recipe DeScoD-ECG on BW), and 75-epoch controls used 2 seeds as exploratory sensitivity checks. Artifact-specific conventional baselines (wavelet shrinkage, median filtering, template subtraction, adaptive notch filtering), simultaneous multi-artifact superpositions, and drop-in (\texttt{raw}-trained, \texttt{denoised}-tested) downstream evaluation remain key priorities for future work.

Third, regarding \emph{generalizability and deployment}, the sEMG sign reversal was evaluated on TD-LDA and TD-SVM across three SNRs on DB2; generality to neural decoders, intermediate SNRs, and other corpora, plus direct within-class variance measurements, requires further study. Finally, all cohorts are retrospective without prospective clinical validation, and single-thread CPU timing measures compute latency rather than end-to-end embedded wearable latency.

\section{Conclusion}

By holding architecture and training recipe fixed across ECG and sEMG benchmarks, we show that model capacity behavior depends strongly on benchmark construction, training-noise diversity, optimization budget, and metric choice. On the standard beat-based ECG benchmark, compact networks achieve competitive reconstruction performance against substantially larger controlled baselines while 40.3K-parameter models degrade under limited training-noise diversity, whereas continuous-window mixed-noise ECG reconstruction improves monotonically to 40.3K parameters. On sEMG, waveform correlation shows diminishing returns beyond 3.2K parameters while SDR and mean-frequency error continue improving for narrowband power-line interference and exhibit an optimization confound for cardiac interference.

Across unmodified clinical ECG and real or contaminated sEMG cohorts, clean-reference reconstruction metrics do not reliably predict task utility: large low-SNR gains under synthetic sEMG contamination contrast with modest, classifier-dependent gains on unmodified clinical ECG and real GRABMyo recordings, and linear sEMG decoding undergoes an SNR-dependent metric--utility sign reversal at $+10$~dB. Future biosignal denoising benchmarks should expand independent noise-source diversity, include compact capacity-controlled baselines, and require downstream task validation alongside reconstruction scores.

\section*{Data and Code Availability}

All datasets analyzed are publicly available: QTDB v1.0.0, MIT-BIH NSTDB v1.0.0, MIT-BIH Arrhythmia Database v1.0.0, MIT-BIH Normal Sinus Rhythm Database v1.0.0, PTB-XL v1.0.3, PhysioNet/CinC Challenge 2017 v1.0.0, CPSC2018 (via PhysioNet/CinC Challenge 2020 v1.0.2), and GRABMyo v1.1.0 via PhysioNet (\url{https://physionet.org/}) \cite{goldberger2000physionet}; NinaPro DB2 via the NinaPro repository (\url{https://ninapro.hevs.ch/}) \cite{atzori2014ninapro}. Mixture-generation scripts, record-level split manifests, training configurations, downstream evaluators, and statistical analysis code will be released publicly upon journal submission.

\section*{Ethics Statement}

This study performs secondary computational analysis of publicly available, de-identified human physiological datasets. Ethical approval and informed consent were obtained by the original dataset curators \cite{laguna1997qtdb,moody2001mitbih,wagner2020ptbxl,clifford2017cinc,liu2018cpsc,atzori2014ninapro,pradhan2022grabmyo}. No new human-subject data were collected.

\section*{Competing Interests and Funding}

The authors declare no competing financial or non-financial interests. This research received no specific grant from funding agencies in the public, commercial, or not-for-profit sectors.

\section*{Acknowledgments}

The authors thank the curators and participants of PhysioNet, PTB-XL, CPSC2018, NinaPro, and GRABMyo for making these physiological recordings openly available, and acknowledge computational resources at the Indian Institute of Technology Mandi.

\bibliographystyle{unsrtnat}
\bibliography{references}

\clearpage
\appendix

\section{Extended Capacity Tables}
\label{app:capacity}

Tables~\ref{tab:app_ecg_fine}, \ref{tab:app_ecg_cont}, and \ref{tab:app_emg_fine} report full per-width reconstruction summaries for beat-based ECG (nv1), continuous-window mixed-noise ECG, and sEMG.

\begin{table}[htbp]
\centering
\caption{Beat-based ECG fine-grid reconstruction performance on QTDB+NSTDB (nv1, mean $\pm$ SD; 5 seeds for base1--base8 and base16, 3 seeds for base10 and base12).}
\label{tab:app_ecg_fine}
\scriptsize
\setlength{\tabcolsep}{4pt}
\begin{tabular}{@{}lrrrrrrr@{}}
\toprule
 & & \multicolumn{2}{c}{\textbf{Baseline Wander (BW)}} & \multicolumn{2}{c}{\textbf{Electrode Motion (EM)}} & \multicolumn{2}{c}{\textbf{Muscle Artifact (MA)}} \\
\cmidrule(lr){3-4}\cmidrule(lr){5-6}\cmidrule(lr){7-8}
\textbf{Width} & \textbf{Params} & \textbf{CC} & \textbf{SDR (dB)} & \textbf{CC} & \textbf{SDR (dB)} & \textbf{CC} & \textbf{SDR (dB)} \\
\midrule
base1 & 392 & $0.618 \pm 0.237$ & $3.18 \pm 2.61$ & $0.577 \pm 0.171$ & $2.43 \pm 1.60$ & $0.685 \pm 0.145$ & $3.43 \pm 2.14$ \\
base2 & 1,048 & $0.793 \pm 0.071$ & $5.20 \pm 1.43$ & $0.770 \pm 0.050$ & $4.66 \pm 0.83$ & $0.811 \pm 0.061$ & $5.41 \pm 1.33$ \\
base3 & 1,990 & $0.880 \pm 0.067$ & $7.54 \pm 2.23$ & $0.773 \pm 0.081$ & $4.69 \pm 1.45$ & $0.878 \pm 0.067$ & $7.40 \pm 2.19$ \\
base4 & 3,218 & $0.898 \pm 0.061$ & $8.15 \pm 2.24$ & $0.816 \pm 0.003$ & $5.40 \pm 0.12$ & $0.924 \pm 0.002$ & $9.07 \pm 0.14$ \\
base5 & 4,732 & $0.929 \pm 0.001$ & $9.42 \pm 0.07$ & $0.816 \pm 0.004$ & $5.41 \pm 0.07$ & $0.927 \pm 0.002$ & $9.30 \pm 0.12$ \\
base6 & 6,532 & $0.927 \pm 0.002$ & $9.30 \pm 0.08$ & $0.815 \pm 0.006$ & $5.27 \pm 0.19$ & $0.928 \pm 0.002$ & $9.38 \pm 0.07$ \\
base8 & 10,990 & $0.928 \pm 0.001$ & $9.40 \pm 0.11$ & $0.812 \pm 0.006$ & $5.12 \pm 0.16$ & $0.927 \pm 0.001$ & $9.35 \pm 0.06$ \\
base10 & 16,592 & $0.929 \pm 0.001$ & $9.48 \pm 0.06$ & $0.799 \pm 0.005$ & $4.69 \pm 0.23$ & $0.927 \pm 0.002$ & $9.42 \pm 0.13$ \\
base12 & 23,338 & $0.928 \pm 0.002$ & $9.44 \pm 0.11$ & $0.796 \pm 0.006$ & $4.53 \pm 0.11$ & $0.923 \pm 0.002$ & $9.29 \pm 0.10$ \\
base16 & 40,262 & $0.925 \pm 0.002$ & $9.36 \pm 0.07$ & $0.774 \pm 0.008$ & $3.77 \pm 0.20$ & $0.917 \pm 0.003$ & $8.98 \pm 0.12$ \\
\bottomrule
\end{tabular}
\end{table}

\begin{table}[htbp]
\centering
\caption{Continuous-window mixed-noise ECG reconstruction performance on the 14 held-out QTDB test records (8,848 windows, mean $\pm$ SD across 3 seeds).}
\label{tab:app_ecg_cont}
\scriptsize
\setlength{\tabcolsep}{4pt}
\begin{tabular}{@{}lrrrrrrr@{}}
\toprule
\textbf{Width} & \textbf{Params} & \textbf{CC} & \textbf{SDR (dB)} & \textbf{T-RRMSE} & \textbf{S-RRMSE} & \textbf{SSD (a.u.)} & \textbf{PRD (\%)} \\
\midrule
base2 & 1,048 & $0.797 \pm 0.028$ & $5.24 \pm 0.71$ & $0.587 \pm 0.041$ & $0.568 \pm 0.031$ & $19.06 \pm 2.59$ & $85.53 \pm 11.44$ \\
base4 & 3,218 & $0.855 \pm 0.007$ & $7.12 \pm 0.17$ & $0.486 \pm 0.010$ & $0.451 \pm 0.016$ & $13.55 \pm 0.67$ & $62.29 \pm 1.50$ \\
base6 & 6,532 & $0.864 \pm 0.005$ & $7.47 \pm 0.15$ & $0.470 \pm 0.007$ & $0.435 \pm 0.009$ & $12.83 \pm 0.34$ & $59.02 \pm 1.70$ \\
base8 & 10,990 & $0.870 \pm 0.004$ & $7.72 \pm 0.08$ & $0.457 \pm 0.006$ & $0.421 \pm 0.005$ & $12.30 \pm 0.38$ & $56.79 \pm 1.03$ \\
base16 & 40,262 & $0.881 \pm 0.001$ & $8.09 \pm 0.04$ & $0.440 \pm 0.002$ & $0.406 \pm 0.008$ & $11.40 \pm 0.12$ & $52.92 \pm 0.77$ \\
\bottomrule
\end{tabular}
\end{table}

\begin{table}[htbp]
\centering
\caption{sEMG extended-grid reconstruction performance on NinaPro DB2 (mean across 5 seeds for base1--base8 and base16, 3 seeds for base10, base12, base24, base32). Matched 3-seed (seeds 42--44) base16 SDR means used for base16$\to$24 and base16$\to$32 paired contrasts in Section~\ref{sec:emg_cap} are 14.35~dB (ecg), 14.53~dB (moa), and 16.41~dB (pli).}
\label{tab:app_emg_fine}
\scriptsize
\setlength{\tabcolsep}{3.5pt}
\begin{tabular}{@{}lrrrrrrrrrr@{}}
\toprule
 & & \multicolumn{3}{c}{\textbf{ECG Interference}} & \multicolumn{3}{c}{\textbf{Motion Artifact (MOA)}} & \multicolumn{3}{c}{\textbf{Power-Line (PLI)}} \\
\cmidrule(lr){3-5}\cmidrule(lr){6-8}\cmidrule(lr){9-11}
\textbf{Width} & \textbf{Params} & \textbf{CC} & \textbf{SDR} & \textbf{MF (Hz)} & \textbf{CC} & \textbf{SDR} & \textbf{MF (Hz)} & \textbf{CC} & \textbf{SDR} & \textbf{MF (Hz)} \\
\midrule
base1 & 392 & 0.843 & 5.83 & 17.55 & 0.837 & 5.78 & 17.00 & 0.874 & 6.44 & 11.31 \\
base2 & 1,048 & 0.907 & 7.93 & 13.85 & 0.897 & 7.75 & 13.53 & 0.906 & 8.32 & 8.89 \\
base3 & 1,990 & 0.950 & 11.33 & 9.74 & 0.940 & 11.43 & 10.45 & 0.956 & 11.67 & 5.50 \\
base4 & 3,218 & 0.969 & 13.38 & 8.61 & 0.963 & 13.84 & 8.98 & 0.978 & 13.88 & 3.62 \\
base5 & 4,732 & 0.971 & 13.83 & 8.23 & 0.964 & 14.16 & 8.67 & 0.979 & 14.26 & 3.55 \\
base6 & 6,532 & 0.971 & 13.81 & 8.42 & 0.964 & 14.09 & 8.74 & 0.981 & 14.59 & 3.28 \\
base8 & 10,990 & 0.970 & 13.87 & 8.40 & 0.965 & 14.32 & 8.34 & 0.983 & 15.13 & 3.15 \\
base10 & 16,592 & 0.969 & 13.79 & 8.22 & 0.965 & 14.39 & 8.37 & 0.984 & 15.58 & 2.87 \\
base12 & 23,338 & 0.972 & 14.28 & 7.32 & 0.966 & 14.46 & 8.20 & 0.985 & 15.85 & 2.71 \\
base16 & 40,262 & 0.972 & 14.32 & 6.95 & 0.966 & 14.52 & 8.17 & 0.986 & 16.28 & 2.66 \\
base24 & 87,838 & 0.973 & 14.28 & 6.44 & 0.965 & 14.50 & 8.19 & 0.989 & 17.19 & 2.38 \\
base32 & 153,718 & 0.972 & 14.14 & 6.51 & 0.965 & 14.54 & 8.31 & 0.990 & 17.66 & 2.24 \\
\bottomrule
\end{tabular}
\end{table}

\section{Optimization Controls and sEMG SNR-Stratified Contrasts}
\label{app:controls}

Table~\ref{tab:app_controls} details the 8.46M-parameter EEGDN CNN recipe-sensitivity control on ECG and the 2-seed 75-epoch sensitivity control on sEMG. Table~\ref{tab:app_emg_snr} reports the sEMG base6$\to$base16 paired increments stratified across all five test SNR levels.

\begin{table}[htbp]
\centering
\caption{Optimization controls: (top) 8.46M-parameter EEGDN CNN recipe sensitivity relative to DSConv base6 (2 seeds, nv1); (bottom) sEMG base16 $-$ base6 SDR differences at 25 vs.\ 75 epochs.}
\label{tab:app_controls}
\scriptsize
\setlength{\tabcolsep}{4pt}
\begin{tabular}{@{}llrrrr@{}}
\toprule
\textbf{Control / Task} & \textbf{Variant} & \textbf{CC} & \textbf{SDR} & \textbf{$\Delta$CC vs.\ b6} & \textbf{$\Delta$SDR vs.\ b6} \\
\midrule
ECG BW (8.46M CNN) & Default (lr $10^{-3}$, 25 ep) & 0.782 & 3.16 & $-0.144$ & $-6.10$ \\
ECG BW (8.46M CNN) & lr $3\times 10^{-4}$, 25 ep & 0.812 & 4.53 & $-0.115$ & $-4.73$ \\
ECG BW (8.46M CNN) & lr $10^{-3}$, 75 ep & 0.742 & 1.56 & $-0.185$ & $-7.69$ \\
ECG EM (8.46M CNN) & lr $3\times 10^{-4}$, 25 ep & 0.674 & 0.12 & $-0.139$ & $-5.19$ \\
ECG MA (8.46M CNN) & lr $3\times 10^{-4}$, 25 ep & 0.832 & 4.73 & $-0.096$ & $-4.61$ \\
\midrule
\textbf{sEMG 75-ep control} & \textbf{Contrast} & \textbf{S42 (25 ep)} & \textbf{S42 (75 ep)} & \textbf{S43 (25 ep)} & \textbf{S43 (75 ep)} \\
\midrule
sEMG ECG interference & base16 $-$ base6 SDR (dB) & $+0.784$ & $-0.006$ & $+0.451$ & $-0.007$ \\
sEMG MOA & base16 $-$ base6 SDR (dB) & $+0.808$ & $+0.175$ & $+0.223$ & $+0.273$ \\
sEMG PLI & base16 $-$ base6 SDR (dB) & $+1.625$ & $+2.297$ & $+1.703$ & $+2.001$ \\
\bottomrule
\end{tabular}
\end{table}

\begin{table}[htbp]
\centering
\caption{sEMG SNR-stratified base6$\to$base16 matched-seed increments (5 seeds, 95\% paired-$t$ CIs) for CC, SDR (dB), and oriented MF-RMSE reduction (Hz, positive = lower error at base16).}
\label{tab:app_emg_snr}
\scriptsize
\setlength{\tabcolsep}{3.5pt}
\begin{tabular}{@{}lrrrr@{}}
\toprule
\textbf{Task} & \textbf{Test SNR} & \textbf{$\Delta$CC [95\% CI]} & \textbf{$\Delta$SDR (dB) [95\% CI]} & \textbf{$\Delta$MF-RMSE (Hz) [95\% CI]} \\
\midrule
ECG & $-14$~dB & $+0.0012$ $[-0.0033, +0.0057]$ & $+0.21$ $[-0.11, +0.54]$ & $+1.22$ $[-0.29, +2.72]$ \\
ECG & $-10$~dB & $+0.0009$ $[-0.0010, +0.0028]$ & $+0.17$ $[-0.08, +0.42]$ & $+1.22$ $[+0.34, +2.11]$ \\
ECG & $-6$~dB & $+0.0011$ $[-0.0002, +0.0023]$ & $+0.28$ $[+0.02, +0.54]$ & $+1.30$ $[+0.64, +1.96]$ \\
ECG & $-2$~dB & $+0.0022$ $[+0.0004, +0.0040]$ & $+0.66$ $[+0.25, +1.07]$ & $+1.62$ $[+0.80, +2.44]$ \\
ECG & $+2$~dB & $+0.0032$ $[+0.0007, +0.0058]$ & $+1.21$ $[+0.57, +1.86]$ & $+2.00$ $[+0.95, +3.04]$ \\
\midrule
MOA & $-14$~dB & $+0.0043$ $[+0.0021, +0.0065]$ & $+0.28$ $[+0.14, +0.43]$ & $+1.03$ $[+0.29, +1.77]$ \\
MOA & $-10$~dB & $+0.0016$ $[+0.0008, +0.0025]$ & $+0.21$ $[+0.10, +0.32]$ & $+0.70$ $[+0.15, +1.24]$ \\
MOA & $-6$~dB & $+0.0009$ $[+0.0003, +0.0015]$ & $+0.29$ $[+0.08, +0.50]$ & $+0.45$ $[+0.01, +0.89]$ \\
MOA & $-2$~dB & $+0.0008$ $[+0.0002, +0.0013]$ & $+0.50$ $[+0.10, +0.90]$ & $+0.33$ $[-0.02, +0.68]$ \\
MOA & $+2$~dB & $+0.0008$ $[+0.0002, +0.0014]$ & $+0.88$ $[+0.10, +1.66]$ & $+0.35$ $[-0.01, +0.71]$ \\
\midrule
PLI & $-14$~dB & $+0.0070$ $[+0.0052, +0.0088]$ & $+2.02$ $[+1.59, +2.45]$ & $+0.68$ $[+0.54, +0.83]$ \\
PLI & $-10$~dB & $+0.0054$ $[+0.0039, +0.0070]$ & $+1.70$ $[+1.30, +2.11]$ & $+0.55$ $[+0.46, +0.64]$ \\
PLI & $-6$~dB & $+0.0053$ $[+0.0032, +0.0074]$ & $+1.63$ $[+1.12, +2.15]$ & $+0.54$ $[+0.45, +0.62]$ \\
PLI & $-2$~dB & $+0.0055$ $[+0.0028, +0.0081]$ & $+1.59$ $[+0.95, +2.23]$ & $+0.60$ $[+0.45, +0.74]$ \\
PLI & $+2$~dB & $+0.0055$ $[+0.0025, +0.0085]$ & $+1.47$ $[+0.78, +2.16]$ & $+0.73$ $[+0.40, +1.05]$ \\
\bottomrule
\end{tabular}
\end{table}

\section{Metric--Utility Small-Cluster Inference Sensitivity}
\label{app:mu}

Tables~\ref{tab:app_ecg_mu} and \ref{tab:app_emg_mu} report complete within-unit slopes, 95\% CR1 confidence intervals, and BH-FDR $q$-values across all five inference methods (CR1, CR2/Satterthwaite, Monte Carlo Webb wild cluster bootstrap, two-stage unit-slope $t$, and two-stage Wilcoxon) for both ECG (30 tests) and sEMG (24 tests).

\begin{table}[htbp]
\centering
\caption{ECG within-record metric--utility slopes (per within-record metric SD) with 95\% CR1 CIs and BH-FDR $q$-values across all five inference methods. MIT-BIH endpoint: R-peak F1 on the independent unseen-parent DS2 subset ($n=15$ records, 675 observations; on the full 22-record DS2 benchmark, 25/30 CR1 and 21/30 CR2 tests satisfy $q < 0.05$). QTDB endpoint: oriented 4-fiducial MAE reduction in ms ($n=14$ records, 630 observations).}
\label{tab:app_ecg_mu}
\scriptsize
\setlength{\tabcolsep}{2.4pt}
\resizebox{\textwidth}{!}{%
\begin{tabular}{@{}lllrlrrrrrr@{}}
\toprule
\textbf{Cohort} & \textbf{Noise} & \textbf{Metric} & \textbf{Slope [95\% CI]} & \textbf{CR1 $t$} & \textbf{CR1 $q$} & \textbf{CR2 $q$} & \textbf{Webb $q$} & \textbf{2S-$t$ $q$} & \textbf{2S-W $q$} & \textbf{Pos.} \\
\midrule
MIT-BIH & BW & CC & $+0.0011$ $[-0.0070, +0.0093]$ & $+0.30$ & 0.795 & 0.815 & 0.799 & 0.713 & 0.599 & 10/15 \\
MIT-BIH & BW & T-RRMSE & $+0.0001$ $[-0.0078, +0.0080]$ & $+0.02$ & 0.986 & 0.986 & 0.986 & 0.713 & 0.599 & 9/15 \\
MIT-BIH & BW & S-RRMSE & $+0.0021$ $[-0.0021, +0.0062]$ & $+1.06$ & 0.353 & 0.388 & 0.470 & 0.133 & 0.195 & 10/15 \\
MIT-BIH & BW & SDR & $+0.0019$ $[-0.0031, +0.0069]$ & $+0.81$ & 0.481 & 0.508 & 0.558 & 0.580 & 0.543 & 10/15 \\
MIT-BIH & BW & $\Delta$SNR & $-0.0021$ $[-0.0109, +0.0068]$ & $-0.51$ & 0.666 & 0.680 & 0.733 & 0.652 & 0.599 & 10/15 \\
MIT-BIH & EM & CC & $+0.0208$ $[+0.0157, +0.0260]$ & $+8.61$ & $<10^{-4}$ & $<0.001$ & 0.002 & $<0.001$ & $<0.001$ & 15/15 \\
MIT-BIH & EM & T-RRMSE & $+0.0193$ $[+0.0138, +0.0248]$ & $+7.57$ & $<10^{-4}$ & $<0.001$ & $<0.001$ & $<0.001$ & $<0.001$ & 15/15 \\
MIT-BIH & EM & S-RRMSE & $+0.0156$ $[+0.0069, +0.0244]$ & $+3.84$ & 0.005 & 0.084 & 0.026 & $<0.001$ & $<0.001$ & 15/15 \\
MIT-BIH & EM & SDR & $+0.0181$ $[+0.0126, +0.0236]$ & $+7.08$ & $<10^{-4}$ & $<0.001$ & $<0.001$ & 0.003 & $<0.001$ & 15/15 \\
MIT-BIH & EM & $\Delta$SNR & $+0.0179$ $[+0.0126, +0.0232]$ & $+7.27$ & $<10^{-4}$ & $<0.001$ & $<0.001$ & $<0.001$ & $<0.001$ & 15/15 \\
MIT-BIH & MA & CC & $+0.0076$ $[+0.0005, +0.0146]$ & $+2.30$ & 0.053 & 0.123 & 0.038 & 0.100 & 0.050 & 13/15 \\
MIT-BIH & MA & T-RRMSE & $+0.0069$ $[+0.0001, +0.0138]$ & $+2.17$ & 0.062 & 0.107 & 0.027 & 0.035 & 0.028 & 13/15 \\
MIT-BIH & MA & S-RRMSE & $+0.0060$ $[+0.0002, +0.0118]$ & $+2.21$ & 0.060 & 0.099 & 0.022 & 0.030 & $<0.001$ & 14/15 \\
MIT-BIH & MA & SDR & $+0.0063$ $[-0.0000, +0.0127]$ & $+2.13$ & 0.064 & 0.103 & 0.022 & 0.030 & 0.006 & 13/15 \\
MIT-BIH & MA & $\Delta$SNR & $+0.0063$ $[-0.0001, +0.0127]$ & $+2.10$ & 0.065 & 0.103 & 0.022 & 0.041 & 0.011 & 13/15 \\
\midrule
QTDB & BW & CC & $+3.55$ $[+0.88, +6.23]$ & $+2.87$ & 0.023 & 0.084 & 0.017 & 0.002 & 0.008 & 13/14 \\
QTDB & BW & T-RRMSE & $+3.37$ $[+0.94, +5.80]$ & $+2.99$ & 0.022 & 0.066 & 0.011 & 0.002 & 0.004 & 13/14 \\
QTDB & BW & S-RRMSE & $+3.48$ $[+1.44, +5.53]$ & $+3.67$ & 0.008 & 0.022 & 0.006 & 0.001 & $<0.001$ & 13/14 \\
QTDB & BW & SDR & $+3.30$ $[+1.13, +5.48]$ & $+3.28$ & 0.015 & 0.042 & 0.006 & 0.003 & 0.002 & 13/14 \\
QTDB & BW & $\Delta$SNR & $+3.27$ $[+1.06, +5.48]$ & $+3.20$ & 0.016 & 0.043 & 0.006 & 0.004 & 0.002 & 13/14 \\
QTDB & EM & CC & $+3.63$ $[+2.35, +4.91]$ & $+6.12$ & $<0.001$ & 0.005 & 0.005 & $<0.001$ & $<0.001$ & 14/14 \\
QTDB & EM & T-RRMSE & $+3.59$ $[+2.41, +4.76]$ & $+6.59$ & $<10^{-4}$ & 0.002 & 0.004 & $<0.001$ & $<0.001$ & 14/14 \\
QTDB & EM & S-RRMSE & $+3.19$ $[+1.83, +4.54]$ & $+5.09$ & $<0.001$ & 0.012 & 0.005 & $<0.001$ & $<0.001$ & 14/14 \\
QTDB & EM & SDR & $+3.63$ $[+2.46, +4.80]$ & $+6.69$ & $<10^{-4}$ & 0.002 & 0.003 & $<0.001$ & $<0.001$ & 14/14 \\
QTDB & EM & $\Delta$SNR & $+3.51$ $[+2.38, +4.64]$ & $+6.73$ & $<10^{-4}$ & 0.002 & 0.003 & $<10^{-4}$ & $<0.001$ & 14/14 \\
QTDB & MA & CC & $+2.98$ $[+0.65, +5.30]$ & $+2.77$ & 0.026 & 0.103 & 0.047 & 0.118 & 0.025 & 12/14 \\
QTDB & MA & T-RRMSE & $+2.86$ $[+0.72, +5.00]$ & $+2.89$ & 0.023 & 0.076 & 0.033 & 0.069 & 0.022 & 12/14 \\
QTDB & MA & S-RRMSE & $+2.82$ $[+0.59, +5.05]$ & $+2.73$ & 0.026 & 0.077 & 0.042 & 0.020 & 0.015 & 12/14 \\
QTDB & MA & SDR & $+2.78$ $[+0.75, +4.80]$ & $+2.97$ & 0.022 & 0.057 & 0.024 & 0.039 & 0.018 & 12/14 \\
QTDB & MA & $\Delta$SNR & $+2.67$ $[+0.57, +4.78]$ & $+2.75$ & 0.026 & 0.066 & 0.027 & 0.039 & 0.035 & 12/14 \\
\bottomrule
\end{tabular}%
}
\end{table}

\begin{table}[htbp]
\centering
\caption{sEMG within-subject metric--utility slopes on NinaPro DB2 ($n=40$ subjects, 600 observations per cell; pp accuracy per within-subject metric SD) with 95\% CR1 CIs and BH-FDR $q$-values across all five inference methods (CR1, CR2/Satterthwaite, Webb wild cluster bootstrap, two-stage $t$, two-stage Wilcoxon). All 24 tests satisfy $q < 0.05$ under every method.}
\label{tab:app_emg_mu}
\scriptsize
\setlength{\tabcolsep}{2.4pt}
\resizebox{\textwidth}{!}{%
\begin{tabular}{@{}lllrlrrrrrr@{}}
\toprule
\textbf{Clf.} & \textbf{SNR} & \textbf{Metric} & \textbf{Slope [95\% CI] (pp)} & \textbf{CR1 $t$} & \textbf{CR1 $q$} & \textbf{CR2 $q$} & \textbf{Webb $q$} & \textbf{2S-$t$ $q$} & \textbf{2S-W $q$} & \textbf{Pos.} \\
\midrule
TD-LDA & $-10$ & CC & $+0.30$ $[+0.17, +0.43]$ & $+4.72$ & $<10^{-4}$ & $<10^{-4}$ & $<10^{-4}$ & $<10^{-4}$ & $<10^{-5}$ & 31/40 \\
TD-LDA & $-10$ & SDR & $+0.20$ $[+0.08, +0.32]$ & $+3.49$ & 0.001 & 0.002 & $<0.001$ & 0.002 & 0.003 & 27/40 \\
TD-LDA & $-10$ & ARV-RMSE & $+0.93$ $[+0.79, +1.07]$ & $+13.73$ & $<10^{-14}$ & $<10^{-13}$ & $<10^{-4}$ & $<10^{-12}$ & $<10^{-9}$ & 39/40 \\
TD-LDA & $-10$ & MF-RMSE & $+0.61$ $[+0.49, +0.73]$ & $+10.26$ & $<10^{-11}$ & $<10^{-10}$ & $<10^{-4}$ & $<10^{-10}$ & $<10^{-10}$ & 39/40 \\
TD-LDA & $0$ & CC & $+0.18$ $[+0.04, +0.32]$ & $+2.66$ & 0.011 & 0.011 & 0.011 & 0.014 & 0.025 & 25/40 \\
TD-LDA & $0$ & SDR & $+0.24$ $[+0.10, +0.38]$ & $+3.51$ & 0.001 & 0.002 & 0.002 & 0.001 & 0.002 & 28/40 \\
TD-LDA & $0$ & ARV-RMSE & $+0.21$ $[+0.08, +0.34]$ & $+3.30$ & 0.002 & 0.002 & 0.002 & 0.022 & 0.023 & 25/40 \\
TD-LDA & $0$ & MF-RMSE & $+0.50$ $[+0.38, +0.61]$ & $+8.69$ & $<10^{-9}$ & $<10^{-8}$ & $<10^{-4}$ & $<10^{-7}$ & $<10^{-6}$ & 35/40 \\
TD-LDA & $+10$ & CC & $-0.38$ $[-0.49, -0.28]$ & $-7.20$ & $<10^{-7}$ & $<10^{-7}$ & $<10^{-4}$ & $<10^{-7}$ & $<10^{-7}$ & 5/40 \\
TD-LDA & $+10$ & SDR & $-0.46$ $[-0.55, -0.37]$ & $-9.96$ & $<10^{-10}$ & $<10^{-10}$ & $<10^{-4}$ & $<10^{-10}$ & $<10^{-9}$ & 2/40 \\
TD-LDA & $+10$ & ARV-RMSE & $-0.37$ $[-0.48, -0.26]$ & $-6.66$ & $<10^{-6}$ & $<10^{-6}$ & $<10^{-4}$ & $<10^{-7}$ & $<10^{-7}$ & 5/40 \\
TD-LDA & $+10$ & MF-RMSE & $-0.36$ $[-0.46, -0.26]$ & $-7.47$ & $<10^{-7}$ & $<10^{-7}$ & $<10^{-4}$ & $<10^{-8}$ & $<10^{-9}$ & 5/40 \\
\midrule
TD-SVM & $-10$ & CC & $+0.59$ $[+0.44, +0.73]$ & $+8.11$ & $<10^{-8}$ & $<10^{-8}$ & $<10^{-4}$ & $<10^{-9}$ & $<10^{-9}$ & 36/40 \\
TD-SVM & $-10$ & SDR & $+0.53$ $[+0.41, +0.65]$ & $+8.90$ & $<10^{-9}$ & $<10^{-9}$ & $<10^{-4}$ & $<10^{-10}$ & $<10^{-9}$ & 38/40 \\
TD-SVM & $-10$ & ARV-RMSE & $+1.13$ $[+0.98, +1.27]$ & $+15.31$ & $<10^{-16}$ & $<10^{-15}$ & $<10^{-4}$ & $<10^{-15}$ & $<10^{-10}$ & 39/40 \\
TD-SVM & $-10$ & MF-RMSE & $+1.12$ $[+0.99, +1.25]$ & $+17.47$ & $<10^{-18}$ & $<10^{-16}$ & $<10^{-4}$ & $<10^{-17}$ & $<10^{-10}$ & 40/40 \\
TD-SVM & $0$ & CC & $+0.58$ $[+0.44, +0.72]$ & $+8.21$ & $<10^{-9}$ & $<10^{-8}$ & $<10^{-4}$ & $<10^{-8}$ & $<10^{-7}$ & 37/40 \\
TD-SVM & $0$ & SDR & $+0.71$ $[+0.56, +0.86]$ & $+9.47$ & $<10^{-10}$ & $<10^{-10}$ & $<10^{-4}$ & $<10^{-10}$ & $<10^{-9}$ & 38/40 \\
TD-SVM & $0$ & ARV-RMSE & $+0.58$ $[+0.44, +0.72]$ & $+8.40$ & $<10^{-9}$ & $<10^{-8}$ & $<10^{-4}$ & $<10^{-9}$ & $<10^{-8}$ & 37/40 \\
TD-SVM & $0$ & MF-RMSE & $+1.10$ $[+0.94, +1.26]$ & $+14.08$ & $<10^{-15}$ & $<10^{-14}$ & $<10^{-4}$ & $<10^{-14}$ & $<10^{-10}$ & 40/40 \\
TD-SVM & $+10$ & CC & $-0.23$ $[-0.36, -0.10]$ & $-3.52$ & 0.001 & 0.001 & 0.001 & $<0.001$ & $<0.001$ & 9/40 \\
TD-SVM & $+10$ & SDR & $-0.29$ $[-0.40, -0.18]$ & $-5.23$ & $<10^{-5}$ & $<10^{-5}$ & $<10^{-4}$ & $<10^{-4}$ & $<10^{-4}$ & 8/40 \\
TD-SVM & $+10$ & ARV-RMSE & $-0.23$ $[-0.36, -0.10]$ & $-3.64$ & $<0.001$ & 0.001 & $<0.001$ & $<0.001$ & $<0.001$ & 10/40 \\
TD-SVM & $+10$ & MF-RMSE & $-0.17$ $[-0.29, -0.05]$ & $-2.83$ & 0.007 & 0.008 & 0.007 & $<0.001$ & $<0.001$ & 11/40 \\
\bottomrule
\end{tabular}%
}
\end{table}

\section{Metric Definitions, Cross-Modality Comparison, and Source Provenance}
\label{app:provenance}

\textbf{Reconstruction metric definitions.} For clean segment $x \in \mathbb{R}^T$ and denoised estimate $\hat{x} \in \mathbb{R}^T$ (zero-mean versions $\bar{x}, \bar{\hat{x}}$; averaged across segments):
\begin{align*}
\mathrm{CC}(x,\hat{x}) &= \frac{\sum_t \bar{x}_t \bar{\hat{x}}_t}{\sqrt{\sum_t \bar{x}_t^2 \sum_t \bar{\hat{x}}_t^2 + \epsilon}}, &
\mathrm{SDR}(x,\hat{x}) &= 10\log_{10}\frac{\|x\|_2^2 + \epsilon}{\|x - \hat{x}\|_2^2 + \epsilon}, \\
\mathrm{T\text{-}RRMSE}(x,\hat{x}) &= \frac{\|x - \hat{x}\|_2}{\|x\|_2 + \epsilon}, &
\mathrm{S\text{-}RRMSE}(x,\hat{x}) &= \frac{\|P_x - P_{\hat{x}}\|_2}{\|P_x\|_2 + \epsilon}, \\
\mathrm{SSD}(x,\hat{x}) &= \sum_{t=1}^{T}(x_t - \hat{x}_t)^2, &
\mathrm{MAD}(x,\hat{x}) &= \max_{1 \le t \le T} |x_t - \hat{x}_t|, \\
\mathrm{COS\text{-}SIM}(x,\hat{x}) &= \frac{\langle x, \hat{x}\rangle}{\|x\|_2\,\|\hat{x}\|_2}, &
\mathrm{PRD}_{\text{DF}}(x,\hat{x}) &= 100\,\frac{\|\hat{x} - x\|_2}{\|\hat{x} - \mu_{\text{all}}\|_2},
\end{align*}
where $P_x, P_{\hat{x}}$ are Welch PSD estimates (\texttt{nperseg} $= 256$), $\mathrm{SNR}_{\text{imp}} = \mathrm{SDR}(x,\hat{x}) - \mathrm{SNR}_{\text{in}}$, and $\mu_{\text{all}}$ is the global mean of all clean test samples. Crucially, ECG $\mathrm{PRD}_{\text{DF}}$ (computed in mV) follows the exact DeepFilter/DeScoD-ECG codebase convention \cite{romero2021deepfilter,li2024descod}, placing the denoised estimate $\|\hat{x} - \mu_{\text{all}}\|_2$ in the denominator rather than $\|x\|_2$; consequently, when a low-capacity model exhibits amplitude shrinkage ($\|\hat{x}\|_2 < \|x\|_2$), $\mathrm{PRD}_{\text{DF}}$ exceeds $100\times \mathrm{T\text{-}RRMSE}$ (e.g., 85.53\% vs.\ $100 \times 0.587$ at base2 in Table~\ref{tab:app_ecg_cont}, narrowing to 52.92\% vs.\ $100 \times 0.440$ at base16). For sEMG (on the recording $\max(|x|)$-normalized scale over $K$ non-overlapping $W=200$-sample windows), PRD follows the clean-reference convention $100\,\|x - \hat{x}\|_2 / \|x\|_2$, $\mathrm{ARV}_k(z) = W^{-1}\sum_{t\in k}|z_t|$, and $\mathrm{MF}_k(z) = (\sum_{f\in[10,500]} f\,|Z_k(f)|) / (\sum_{f\in[10,500]} |Z_k(f)| + \epsilon)$ using Hann-windowed rFFT magnitudes $Z_k(f)$, with $\mathrm{ARV\text{-}RMSE}$ and $\mathrm{MF\text{-}RMSE}$ computed as the RMSE across windows $k=1,\dots,K$.

\textbf{Protocol provenance.} Primary ECG and sEMG contrasts, margins, and BH-FDR families were prespecified. Fine-grid ECG widths (base1/3/5), extended sEMG widths (base24/32), large-CNN controls, sEMG 1D-CNN decoding, sEMG metric--utility slopes, and real-recording GRABMyo evaluation were secondary exploratory analyses; GRABMyo non-inferiority ($-1$~pp) and primary widths (base6/16) were fixed before evaluation.

\textbf{QTDB split and MIT-BIH parent-record provenance.} The 14 QTDB test records are \texttt{sel123}, \texttt{sel233}, \texttt{sel302}, \texttt{sel307}, \texttt{sel820}, \texttt{sel853}, \texttt{sel16420}, \texttt{sel16795}, \texttt{sele0106}, \texttt{sele0121}, \texttt{sel32}, \texttt{sel49}, \texttt{sel14046}, and \texttt{sel15814}. The remaining 91 QTDB records form the non-test pool (all 91 contribute beats to the beat-based benchmark; split 68 train / 23 validation in the continuous-window benchmark). Fifteen QTDB records originate from MIT-BIH Arrhythmia recordings: \texttt{sel123} and \texttt{sel233} are in QTDB test; \texttt{sel100}, \texttt{sel102}, \texttt{sel103}, \texttt{sel104}, \texttt{sel114}, \texttt{sel116}, \texttt{sel117}, \texttt{sel213}, \texttt{sel223}, and \texttt{sel232} are in continuous-window training; and \texttt{sel221}, \texttt{sel230}, and \texttt{sel231} are in continuous-window validation. Accordingly, 15 of 22 MIT-BIH DS2 records (\texttt{105}, \texttt{111}, \texttt{113}, \texttt{121}, \texttt{123}, \texttt{200}, \texttt{202}, \texttt{210}, \texttt{212}, \texttt{214}, \texttt{219}, \texttt{222}, \texttt{228}, \texttt{233}, \texttt{234}) have zero parent-record overlap with QTDB denoiser training or validation.

\begin{table}[htbp]
\centering
\caption{Cross-modality summary contrasting our prior EEG study \cite{bindra2026eeg} with the new ECG and sEMG results obtained under the same controlled DSConv-ECA backbone.}
\label{tab:app_crossmod}
\scriptsize
\setlength{\tabcolsep}{3.5pt}
\begin{tabularx}{\textwidth}{@{} >{\raggedright\arraybackslash}p{2.3cm} >{\raggedright\arraybackslash}X >{\raggedright\arraybackslash}X >{\raggedright\arraybackslash}X @{}}
\toprule
\textbf{Dimension} & \textbf{EEG (Prior study \cite{bindra2026eeg})} & \textbf{ECG (This study)} & \textbf{sEMG (This study)} \\
\midrule
Signal structure & Non-stationary oscillatory rhythms; spectral overlap with EOG/EMG & Quasi-periodic, low-dimensional P--QRS--T morphology & Broadband stochastic interference pattern \\
\addlinespace
Benchmark capacity regime & EMG-EEG CC saturates by base6--base8 (6.5--11K); EOG gains slightly to base16 & Beat-based peaks at base4--base6 (3.2--6.5K) and declines at base16; continuous-window gains monotonically to base16 & CC shows diminishing returns beyond base4 (3.2K); PLI SDR and MF-RMSE keep gaining to base32 (153.7K) \\
\addlinespace
Dominant confound identified & Fixed source pools (EEGDenoiseNet) & Limited 634-window NSTDB training noise diversity on beat-based split & Optimization budget confound at 25 epochs (ECG-in-EMG SDR gap vanishes at 75 epochs) \\
\addlinespace
Downstream utility (real/matched) & Degrades CSP+LDA motor imagery and Sleep-EDF; mixed/neutral on neural decoders & Small deterministic gains (PTB-XL, CinC); no supported gain on \texttt{xresnet1d101} or CPSC2018 & Large low-SNR DB2 gains (LDA, SVM, 1D-CNN); non-inferior on real GRABMyo (+0.6--0.7~pp over conv.\ filter) \\
\addlinespace
Metric--utility link & Conditional (present in some EMG/mixed conditions, inconsistent for EOG) & Positive in 20/30 CR1 (25/30 Webb) on independent MIT-BIH + QTDB; null on MIT-BIH BW & \textbf{SNR-dependent sign reversal} for linear decoders: positive at $-10/0$~dB, negative at $+10$~dB \\
\bottomrule
\end{tabularx}
\end{table}

\end{document}